\documentstyle[12pt,epsfig]{article}
\begin{document}

\newcommand\beq{\begin{equation}}
\newcommand\eeq{\end{equation}}
\newcommand\bea{\begin{eqnarray}}
\newcommand\eea{\end{eqnarray}}
\newcommand\nonu{\nonumber}
\newcommand\jp{J^\prime }
\newcommand\tphi{{\tilde \phi}}

\begin{center}
\Large {\bf Field Theoretic Studies of Quantum Spin Systems in One Dimension}
\end{center}

\vskip 1 true cm
\centerline{Diptiman Sen \footnote{E-mail: ~diptiman@cts.iisc.ernet.in}}

\vskip .5 true cm
\centerline{\it Centre for Theoretical Studies,}
\centerline{\it Indian Institute of Science, Bangalore 560012} 

\vskip 1 true cm
\begin{abstract}

We describe some field theoretic methods for studying quantum spin systems in 
one dimension. These include the nonlinear $\sigma$-model approach which is 
particularly useful for large values of the spin, the idea of Luttinger liquids
and bosonization which are more useful for small values of spin such as 
spin-$1/2$, and the technique of low-energy effective Hamiltonians which can be
useful if the system under consideration is perturbatively close to an exactly
solvable model. We apply these techniques to similar spin models, such as spin
chains with dimerization and frustration, and spin ladders in the presence of 
a magnetic field. This comparative study illustrates the relative strengths of
the different methods.

\end{abstract}
\vskip 1.5 true cm

\section{Introduction}

One-dimensional and quasi-one-dimensional quantum spin systems have been
studied extensively in recent years for several reasons. Many such systems
have been realized experimentally, and a variety of theoretical techniques, 
both analytical and numerical, are available to study the relevant
models. Due to large quantum fluctuations in low dimensions, 
such systems often have unusual properties such as a gap between a singlet 
ground state and the excited nonsinglet states; this leads to a 
magnetic susceptibility which vanishes exponentially at low temperatures.
Perhaps the most famous example of this is the Haldane gap which was
predicted theoretically in integer spin Heisenberg antiferromagnetic
chains \cite{hald1}, and then observed experimentally in a spin-$1$ system
$Ni(C_2 H_8 N_2 )_2 NO_2 (Cl O_4 )$ \cite{buye}. Other examples include the
spin ladder systems in which a small number of one-dimensional spin-$1/2$
chains interact amongst each other \cite{dago}. It has been observed that 
if the number of chains is even, i.e., if each rung of the ladder (which is
the unit cell for the system) contains an even number
of spin-$1/2$ sites, then the system effectively behaves like an integer spin
chain with a gap in the low-energy spectrum. Some two-chain ladders which 
show a gap are $(VO)_2 P_2 O_7$ \cite{eccl}, $Sr Cu_2 O_3$ \cite{azum}
and $Cu_2 (C_5 H_{12} N_2 )_2 Cl_4$ \cite{chab1}. Conversely, a three-chain
ladder which effectively behaves like a half-odd-integer spin chain and 
does {\it not} exhibit a gap is $Sr_2 Cu_3 O_5$ \cite{azum}. A related
observation is that some quasi-one-dimensional systems such as $CuGeO_3$ 
spontaneously dimerize below a spin-Peierls transition temperature 
\cite{hase}; then the unit cell contains two spin-$1/2$ sites and the 
system is gapped. Another interesting class of systems are the alternating
spin chains such as bimetallic molecular magnets. An example is
$NiCu(pbaOH)(H_2 O)_3 \cdot 2H_2 O$ in which spin-$1$'s ($Ni^{2+}$) and 
spin-$1/2$'s ($Cu^{2+}$) alternate. The ground state of these systems have
a nonzero total spin $S_0$. It turns out that there is a gap to states with
spin greater than $S_0$, but no gap to states with spin less than $S_0$.

The results for gaps quoted above are all in the absence of an external 
magnetic field. The situation becomes even more interesting in the 
presence of a magnetic field \cite{chab2}. Then it is possible for an 
integer spin chain to be gapless and a half-odd-integer spin chain to show 
a gap above the ground state for appropriate values of the field 
\cite{oshi,cabr,tots1,tone,saka}. This has
been demonstrated in several models using a variety of methods such as exact
diagonalization of small systems, bosonization and conformal field theory
\cite{schu1,affl1}, and perturbation theory \cite{reig}. In particular, it has 
been shown that the magnetization of some systems can exhibit plateaus at 
certain nonzero values for some finite ranges of the magnetic field. 

The plan of this paper is as follows.
In Sec. 2, we discuss the low-energy properties of the dimerized and 
frustrated antiferromagnetic spin chain. In Secs. 3 and 4, we present 
some field theoretic methods which can be used for studying spin chains and 
ladders with or without an external magnetic field \cite{frad}. 
These methods rely on the idea that the low-energy and long-wavelength 
modes of a system (i.e., wavelengths much longer than the lattice spacing $a$ 
if the system is defined on a lattice at the microscopic level) can often be 
described by a continuum field theory. In Sec. 3, we discuss the nonlinear
$\sigma$-model approach, while in Sec. 4, we discuss the concepts of 
Tomonaga-Luttinger liquids and bosonization. In Sec. 5, we discuss the 
low-energy effective Hamiltonian approach and show how it can be 
combined with bosonization to gain an understanding of the magnetic properties
of one-dimensional spin systems.

\section{Spin Chain with Dimerization and Frustration}

Experimental studies of some of the quasi-one-dimensional spin systems have 
shown that besides the nearest neighbor antiferromagnetic exchange, there also 
exists a second neighbor exchange $J_2$ of the same sign and comparable 
magnitude. Such a second neighbor interaction has the effect of frustrating 
the spin alignment favored by the nearest neighbor interaction. Therefore, a 
realistic study of one-dimensional systems requires a model with both 
frustration ($J_2$) and dimerization (governed by a parameter $\delta$).

The Hamiltonian for the frustrated and dimerized antiferromagnetic
spin chain can be written as
\beq
H ~=~ \sum_i ~[ ~1 ~-~ (-1)^i ~\delta ~] ~ {\vec S}_i \cdot 
{\vec S}_{i+1} ~ + ~J_2 ~\sum_i ~{\vec S}_i \cdot {\vec S}_{i+2} ~, 
\label{ham1}
\eeq
where the limits of the summation depend on the boundary condition
(open or periodic). (We have set the average nearest neighbor 
interaction $J_1$ to be equal to $1$ for convenience).
The interactions are schematically shown in Fig. 1. The region of interest is 
defined by $J_2 ~\ge 0$ and $0 \le \delta \le 1$. 

\vspace*{0.2cm}
\begin{figure}[htb]
\begin{center}
\vspace*{-0.4cm}
\epsfig{figure=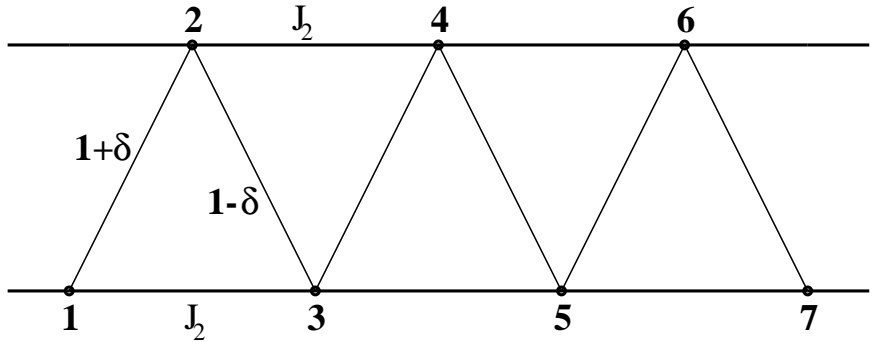,angle=270,width=12cm}
\end{center}
\caption{Schematic picture of the spin chain described by Eq. (\ref{ham1}).}
\label{fig1}
\end{figure}

The ground state properties of the Hamiltonian (\ref{ham1}) have been studied
at some representative points in the $J_2 - \delta$ plane
using the density-matrix renormalization group (DMRG) method \cite{whit}.
The phase diagrams obtained for spin-$1/2$ and spin-$1$ chains are shown in 
Figs. 2 and 3 \cite{pati}. We use the word `phase' only for convenience 
to distinguish between regions with different modulations of the two-spin 
correlation function as discussed later. Our model actually has no phase 
transition even at zero temperature.

For the spin-$1/2$ chain \cite{chit,hamm}, the system is found to be gapless 
on the line $A$ which runs from $J_2 =0$ to $J_{2c} =0.241$ for $\delta =0$; 
see Fig. 2. The model is gapped everywhere else in the $J_2 - \delta$ plane. 
There is a disorder line $B$ given by $2J_2 
+\delta =1$ on which the exact ground state of the model is given by a product
of singlets formed by the nearest-neighbor spins which are joined by the 
stronger bonds ($1+\delta$); this is called the Shastry-Sutherland line 
\cite{shas}, and it ends at the Majumdar-Ghosh point $(J_2 =0.5, \delta =0)$.
The correlation length $\xi$ goes through a minimum on $B$.
Finally, the peak in the structure factor $S(q)$ is at $q_{max} =\pi$ to the 
left of $B$ (called region I), decreases from $\pi$ to $\pi /2$ as one goes 
from $B$ up to the line $C$ (region II), and is at $q_{max} = \pi/2$ to the 
right of $C$ (region III). 

\begin{figure}[htb]
\begin{center}
\epsfig{figure=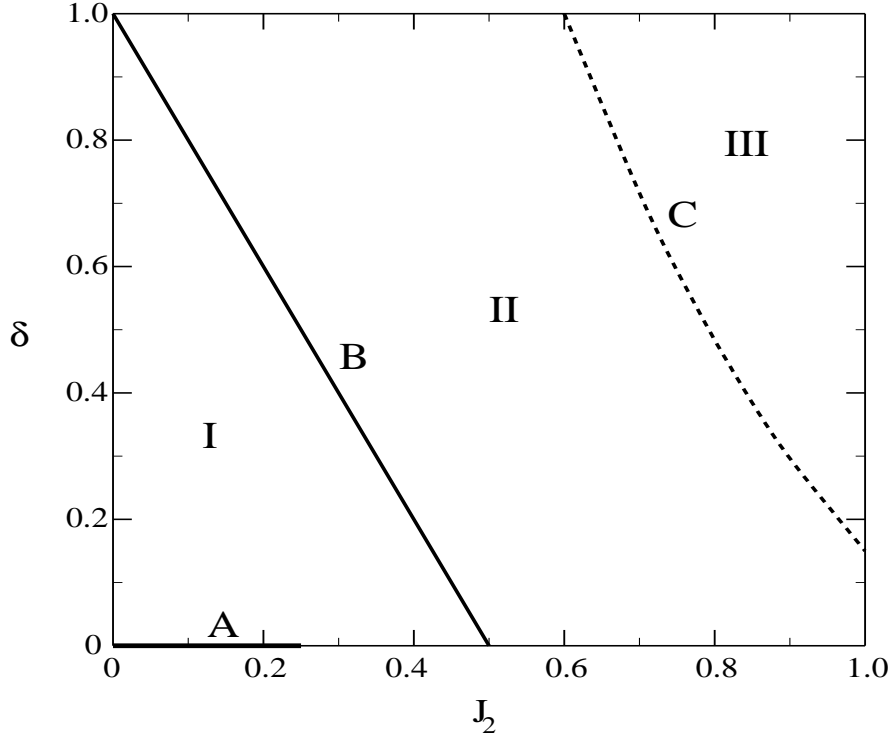,height=10cm,width=12cm}
\end{center}
\caption{Ground state phase diagram of the spin-$1/2$ chain in the $J_2 -
\delta$ plane.}
\label{fig2}
\end{figure}

In the spin-$1$ case (Fig. 3), the phase diagram is more complex. There is a 
solid line marked $A$ which runs from $(0,0.25)$ to about $(0.22 \pm 0.02, 
0.20 \pm 0.02)$ shown by a cross. To within numerical accuracy, the gap is 
zero on this line and the correlation length $\xi$ is as large as the system 
size $N$. The rest of the `phase' diagram 
is gapped. However the gapped portion 
can be divided into different regions characterized by other interesting 
features. On the dotted lines marked $B$, the gap is finite. Although $\xi$ 
goes through a maximum when we cross $B$ in going from region II to region I 
or from region III to region IV, its value is much smaller than $N$. There is 
a dashed line $C$ extending from $(0.65,0.05)$ to about $(0.73,0)$ on which 
the gap appears to be zero (to numerical accuracy), and $\xi$ is very large 
but not as large as $N$. In regions II and III, the ground state for an {\it 
open} chain has a four-fold degeneracy (consisting of $S=0$ and $S=1$), 
whereas it is nondegenerate in regions I and IV with $S=0$. The regions II and
III, where the ground state of an open chain is four-fold degenerate, can be 
identified with the Haldane phase. The regions I and IV correspond to the 
non-Haldane singlet phase. Regions I and IV are separated by the disorder line
$D$ given by $2J_2 + \delta =1$, while regions II and III are separated by
line $E$. The lines $B$, $D$ and $E$ seem to meet in a small region V
where the ground state of the model is numerically very difficult to find.

\begin{figure}[htb]
\begin{center}
\epsfig{figure=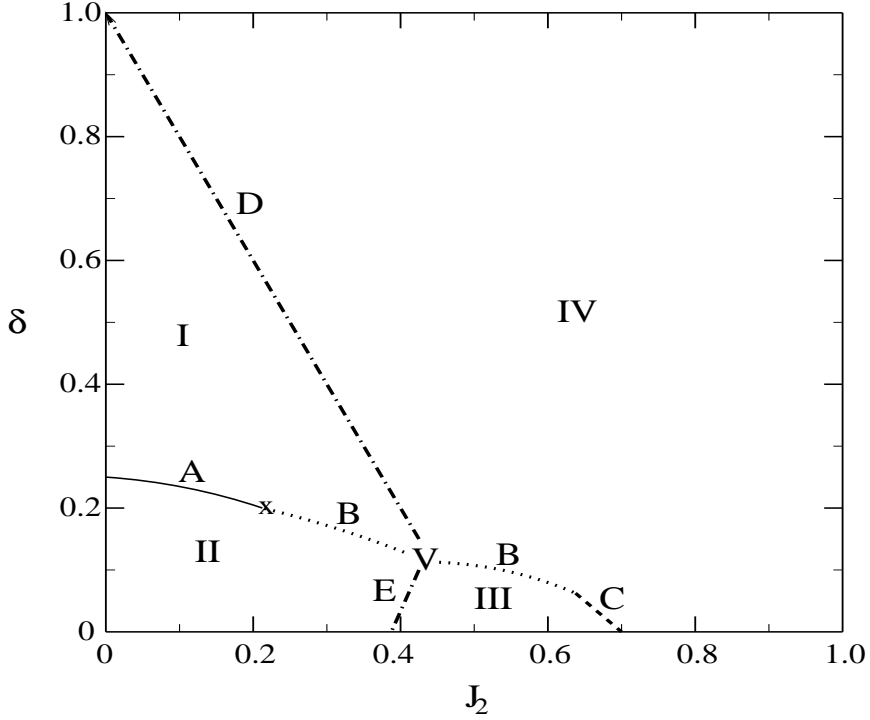,height=10cm,width=12cm}
\end{center}
\caption{Ground state phase diagram of the spin-$1$ chain.}
\label{fig3}
\end{figure}

As can be seen from Fig. 1, setting $\delta =1$ results in a two-chain ladder 
where the interchain coupling is $2$ and the intrachain coupling is $J_2$. 
We can hold $J_2$ fixed and vary the interchain coupling $J$. Numerical 
studies show that for spin-$1/2$, the system is gapped for any nonzero value 
of $J$, although the gap vanishes linearly as $J \rightarrow 0$; this can be
shown using bosonization. On the other hand, the spin-$1$ chain has a finite 
value of the gap for any value of $J$ \cite{pati}.

\section{\bf Nonlinear $\sigma$-model}

The nonlinear $\sigma$-model (NLSM) 
analysis of spin chains with the inclusion of $J_2 ~$ 
and $\delta$ proceeds as follows \cite{rao1}. We first do a classical analysis 
in the $S \rightarrow \infty$ limit to find the ground state configuration of 
the spins. Let us make the ansatz that the ground state is a coplanar 
configuration of the spins with the energy per spin being equal to
\beq
e_0 ~=~ S^2 ~\left[ {1 \over 2} ~ (1+\delta) \cos \theta_1 ~+~
{1 \over 2} ~ (1-\delta ) \cos \theta_2 ~+~ J_2 \cos (\theta_1
+ \theta_2) \right]~, 
\eeq
where $\theta_1$ is the angle between the spins ${\vec S}_{2i}$
and ${\vec S}_{2i+1}$ and $\theta_2$ is the angle between the
spins ${\vec S}_{2i}$ and ${\vec S}_{2i-1}$. Minimization of the classical 
energy with respect to the $\theta_i ~$ yields the following three phases.

\noindent (i) Neel: This phase has $\theta_1 = \theta_2 = \pi$; hence all the
spins point along the same line and they go as $\uparrow \downarrow \uparrow 
\downarrow$ along the chain. This phase is stable for $1-\delta^2 > 4J_2$. 

\noindent (ii) Spiral: Here, the angles $\theta_1$ and $\theta_2$ are given by
\bea
\cos \theta_1 &=& - {1 \over {1+\delta}}~ {\left[~ {{1 - \delta^2}
\over {4 J_2}} ~+~ {\delta \over {1 + \delta^2}}~ 4 J_2 ~ \right]} \nonu \\ 
{\rm and} \quad \cos \theta_2 &=& - {1 \over {1 -\delta}}~ {\left[~ {{1 -
\delta^2} \over {4 J_2}} ~-~ {\delta \over {1 - \delta^2}}~ {4 J_2} ~\right]},
\eea
where $\pi/2 < \theta_1 <\pi$ and $0<\theta_2 < \theta_1$. Thus the spins lie 
on a plane. This phase is stable for $1-\delta^2 < 4 J_2 < (1-\delta^2) /
\delta$.

\noindent (iii) Colinear: This phase (which needs both dimerization and 
frustration) is defined to have $\theta_1 = \pi$ and $\theta_2 = 0$; hence all
the spins point along the same line and they go as $\uparrow \uparrow 
\downarrow \downarrow$ along the chain. It is stable for $(1-\delta^2) /
\delta < 4 J_2$. 

\noindent These phases along with their boundaries are depicted in Fig. 4.
Thus even in the classical limit $S \rightarrow \infty$, the system has
a rich ground state `phase diagram'.

\begin{figure}[htb]
\begin{center}
\epsfig{figure=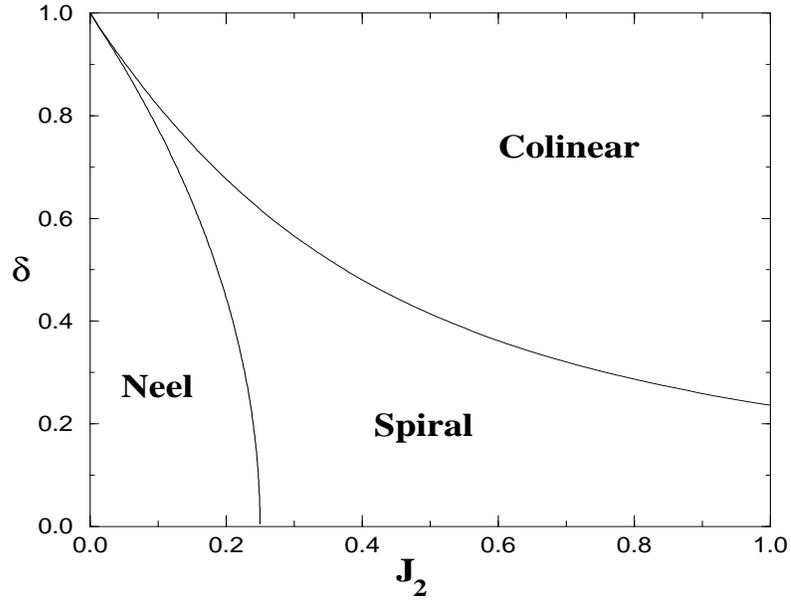,height=9cm,width=12cm}
\end{center}
\vspace*{-0.3cm}
\caption{Classical ground state phase diagram of the spin chain with 
frustration and dimerization.}
\label{fig4}
\end{figure}

We can go to the next order in $1/S$, and
study the spin wave spectrum about the ground state in each of the phases. 
The main results are as follows. In the Neel phase, we find two zero modes, 
i.e., modes for which the energy $\omega_k$ vanishes linearly at certain 
values of the momentum $k$, with the slope $d\omega_k /dk$ at those points
being called the velocity. The two modes are found to have the same velocity
in this phase. In the spiral phase, we have three zero modes, two with the 
same velocity describing out-of-plane fluctuations, and one with a higher 
velocity describing in-plane fluctuations. In the colinear phase, we get two 
zero modes with equal velocities just as in the Neel phase. The three phases 
also differ in the behavior of the spin-spin correlation function $S(q) = 
\sum_n \langle {\vec S}_0 \cdot {\vec S}_n \rangle \exp (-iqn)$ in the 
classical limit. $S(q)$ is peaked at $q = (\theta_1 + \theta_2 )/2$, {\it 
i.e.}, at $q=\pi$ in the Neel phase, at $\pi/2 < q < \pi$ in the spiral phase
and at $q=\pi/2$ in the colinear phase. Even for $S=1/2$ and $1$, DMRG studies
have seen this feature of $S(q)$ in the Neel and spiral phases \cite{pati}. 

\vspace*{0.2cm}
\begin{figure}[htb]
\begin{center}
\vspace*{-0.4cm}
\epsfig{figure=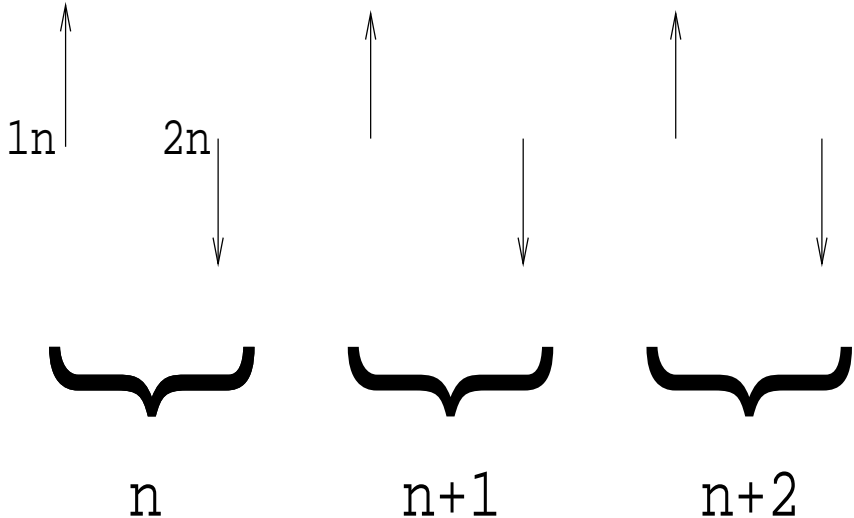,height=8cm,width=12cm}
\vspace*{-0.4cm}
\end{center}
\caption{Classical configuration of the spins in the Neel phase.}
\label{fig5}
\end{figure}

We now derive a NLSM field theory which can describe the low-energy and 
long-wavelength excitations. In the Neel phase, this is given by a $O(3)$ 
NLSM with a topological term \cite{hald1,affl1}. The field variable is a unit 
vector $\vec \phi$ which is defined as follows. The classical ground
state in the Neel phase has a unit cell, labeled by an integer $n$, with 
two sites labeled as $1n$ and $2n$ respectively; see Fig. 5. We define 
linear combinations of the two spins as
\bea
{\vec \phi}_n ~&=&~ \frac{{\vec S}_{1n} ~-~ {\vec S}_{2n}}{2S} ~, \nonu \\
{\vec l}_n ~&=&~ \frac{{\vec S}_{1n} ~+~ {\vec S}_{2n}}{2a} ~.
\label{phil1}
\eea
Here $a$ is the lattice spacing; hence, the size of each unit cell is $2a$.
Note that
\bea
{\vec l}_n ~\cdot ~{\vec \phi}_n ~&=&~ 0 ~, \nonu \\
{\vec \phi}_n^2 ~&=&~ 1 ~+~ \frac{1}{S} ~-~ \frac{a^2 {\vec l}_n^2}{S^2} ~,
\label{const}
\eea
so that ${\vec \phi}_n$ becomes an unit vector in the large $S$ limit. These
fields satisfy the commutation relations
\beq
[ ~l_{ma} ~,~ \phi_{nb} ~]~ = ~\frac{i}{2a}~ \delta_{mn} ~\sum_c ~
\epsilon_{abc} ~\phi_{nc} ~,
\label{comm1}
\eeq
where $m,n$ are unit cell labels, $a,b,c$ denote the components $x,y,z$, and
$\epsilon_{abc}$ is the completely antisymmetric tensor with $\epsilon_{xyz}
=1$. This means that we can write ${\vec l}_n = {\vec \phi}_n \times {\vec 
\Pi}_n$, where the vector $\vec \Pi$ is canonically conjugate to 
$\vec \phi$, i.e.,
\beq
[ ~\phi_{ma} ~,~ \Pi_{nb} ~]~ = ~\frac{i}{2a}~ \delta_{mn} ~\delta_{ab} ~.
\label{comm2}
\eeq
We now go to the continuum limit by introducing a spatial coordinate $x$ which
is equal to $2na$ at the location of the $n^{\rm th}$ unit cell. Summations
get replaced by integrals, i.e., $\sum_n \rightarrow \int dx/(2a)$. 
The commutation relation (\ref{comm2}) then takes the form
\beq
[ ~\phi_a (x)~,~ \Pi_b (y)~]~ = ~i~ \delta (x-y) ~\delta_{ab} ~.
\label{comm3}
\eeq
We note that $\dot {\vec \phi}$ and ${\vec \phi}^{\prime}$ are orthogonal to 
$\vec \phi$ because $\vec \phi$ is an unit vector. 
We will see below that both $\vec l$ and $\vec \Pi$ are given by first-order
space-time derivatives of $\phi$. In the low-energy and long-wavelength
limit, the dominant terms in the Hamiltonian will be those which have 
second-order derivatives of $\vec \phi$, and therefore first-order derivatives
of $\vec l$. To find this Hamiltonian, we rewrite (\ref{ham1}) in terms of
$\vec \phi$ and $\vec l$, and Taylor expand these fields to the necessary 
order, i.e.,
\bea
{\vec \phi}_{n+1} ~&=&~ {\vec \phi} (x) ~+~ 2a {\vec \phi}^{\prime} (x) ~+~ 
2a^2 {\vec \phi}^{\prime \prime} ~+~ \ldots ~, \nonu \\
{\vec l}_{n+1} ~&=&~ {\vec l} (x) ~+~ 2a {\vec l}^{\prime} (x) ~+~ \ldots ~,
\eea
where $x=2na$. We then use the constraints in (\ref{const}) and do some
integration by parts (throwing away boundary terms at $x = \pm \infty$) to
obtain the continuum Hamiltonian
\beq
H ~=~ \int ~dx ~[~ \frac{cg^2}{2} ~({\vec l} + \frac{\theta}{4\pi} 
{\vec \phi}^{\prime})^2 ~ +~ \frac{c}{2g^2} ~{\vec \phi}^{\prime 2} ~]~, 
\label{ham2}
\eeq
where 
\bea
c ~&=&~ 2aS {\sqrt {1 - 4J_2 - \delta^2}} ~, \nonu \\
g^2 ~&=&~ \frac{2}{S{\sqrt {1 - 4 J_2 - \delta^2}}} ~, \nonu \\
{\rm and} \quad \theta ~&=&~ 2 \pi S (1 - \delta) ~.
\label{cc}
\eea
By expanding (\ref{ham2}) to second order in small fluctuations around, 
say, ${\vec \phi} = (0,0,1)$, we find an energy-momentum dispersion relation
of the `massless relativistic' form $\omega = c |k|$; thus $c$ is the
spin wave velocity. Similarly, by expanding (\ref{ham2}) to fourth and higher 
orders in small fluctuations, we find that $g^2$ is the coupling constant 
governing the strength of the interactions between the spin waves. 

One can show that the Hamiltonian (\ref{ham2}) follows from the Lagrangian 
density
\beq
{\cal L} ~=~ {1 \over {2 c g^2}} ~{\dot {\vec \phi}}^2 ~-~ {c \over {2 
g^2}} ~{\vec \phi}^{\prime 2} ~+~ {\theta \over {4 \pi}} ~{\vec \phi} 
\cdot {\vec \phi}^{\prime} \times {\dot {\vec \phi}} ~,
\label{lag1}
\eeq
(Incidentally, one can derive the canonically conjugate momentum $\vec 
\Pi$ and then the angular momentum $\vec l$ from (\ref{lag1}),
\bea
{\vec \Pi} ~&=&~ \frac{1}{cg^2} ~{\dot {\vec \phi}} ~+~ \frac{\theta}{4\pi} ~
{\vec \phi} \times {\vec \phi}^{\prime} ~, \nonu \\
{\vec \l} ~&=&~ {\vec \phi} \times {\vec \Pi} ~=~ \frac{1}{cg^2} ~{\vec \phi} 
\times \dot {\vec \phi} ~-~ \frac{\theta}{4\pi} ~{\vec \phi}^{\prime} ~,
\eea
thereby verifying that $\vec l$ and $\vec \Pi$ only contain first-order
derivatives of $\vec \phi$ as stated above). From (\ref{lag1}),
we see that $\theta$ is the coefficient of a topological term, because the 
integral of this term is an integer which defines the winding number of a 
field configuration ${\vec \phi} (x,t)$). For $\theta = \pi$ mod $2 \pi$ and 
$g$ less than a critical value $g_c$, it is known that
the system is gapless and is described by a conformal field theory with an 
$SU(2)$ symmetry \cite{affl1,affl2}. For any other value of $\theta$, the 
system is gapped, and the gap is of order $\Delta E \sim \exp (-2\pi /g^2)$. 
For $J_2 = \delta =0$, one therefore expects that integer 
spin chains should have a gap of the order $\exp (-\pi S)$ (note that this
goes to zero rapidly as $S \rightarrow \infty$, so that there is no difference
between integer and half-odd-integer spin chains in the classical limit),
while half-integer spin chains should be gapless. For the two-spin equal-time
correlation function, this means that $< {\vec S}_0 \cdot {\vec S}_n >$ 
should decay as a power-law $(-1)^n /|n|$ as $|n| \rightarrow \infty$ for 
half-odd-integer spin chains, and exponentially as $(-1)^n \exp (-n/\xi)$
for integer spin chains, where the correlation length $\xi \sim c/\Delta E$.
All this is known to be true even for small values of $S$ like 
$1/2$ (analytically) and $1$ (numerically) although the field theory is 
only derived for large $S$. In the presence of dimerization, one expects 
a gapless system at certain special values of $\delta$. For $S=1$, the 
special value is predicted to be $\delta_c =0.5$. We see that the {\it 
existence} of a gapless point is correctly predicted by the NLSM. However, 
according to the DMRG results, $\delta_c~$ is at $0.25$ for $J_2 =0$ 
\cite{tots2} and it decreases with $J_2 ~$ as shown in Fig. 3; this differs
from the NLSM results in (\ref{cc}) according to which $\theta$ should be
independent of $J_2$. These deviations from field theory are probably due to 
higher order corrections in $1/S$ which have not been studied analytically 
so far.
 
In the spiral phase of the $J_2 - \delta$ model, it is necessary to use a 
different NLSM which is known for $\delta =0$ \cite{rao2,alle}. The field 
variable is now an $SO(3)$ matrix $\underline R$. The Lagrangian density is
\beq
{\cal L} ~=~ {1 \over {2 c g^2}} ~{\rm tr} ~\Bigl( ~{\dot {\underline R}^T} 
{\dot {\underline R}} ~P_0 ~\Bigr) ~-~ {c \over {2 g^2}} ~{\rm tr} ~\Bigl( ~
{\underline R}^{\prime T} {\underline R}^{\prime} ~P_1 ~\Bigr) ,
\eeq
where $c = S (1 + y) {\sqrt {1 - y^2}} / y$, $g^2 = 2 
{\sqrt {(1 + y) /(1 - y)}} /S$ with $1/y = 4 J_2 ~$, and $P_0 ~$ and 
$P_1 ~$ are diagonal matrices with diagonal elements $(1,1,2 y (1 -
y) / (2 y^2 - 2 y + 1) )$ and $(1,1,0)$ respectively. Note that
there is no topological term; indeed, no such term is possible since $\Pi_2 
(SO(3)) =0$ unlike $\Pi_2 (S^2) = Z$ for the $O(3)$ NLSM in the Neel phase. 
Hence there is no apparent difference between integer and half-integer 
spin chains in the spiral phase. A one-loop renormalization group \cite{rao2} 
and large $N$ analysis \cite{alle} indicate that the system should have a gap 
for all values of $J_2 ~$ and $S$, and that there is no reason for a 
particularly small gap at any special value of $J_2 ~$. The `gapless' point 
found numerically at $J_2 =0.73$ for spin-$1$ is therefore a surprise. 

Finally, in the colinear phase of the $J_2 - \delta$ model, the NLSM is 
known for $\delta =1$, i.e., for the spin ladder. The Lagrangian is the same 
as in (\ref{lag1}), but with $c= 4aS {\sqrt {J_2 (J_2 + 1)}}$, $g^2 = {\sqrt
{1 + 1/J_2}}/S$ and $\theta =0$. There is {\it no} topological term for any 
value of $S$, and the model is therefore gapped.

The field theories for general $\delta$ in both the spiral and 
colinear phases are still not known. Although the results are 
qualitatively expected to be similar to the $\delta=0$ case in the spiral
phase and the $\delta=1$ case in the colinear phase, quantitative features
such as the dependence of the gap on the coupling strengths require the
explicit form of the field theory.

The NLSMs derived above can be expected to be accurate only for large values 
of the spin $S$. It is interesting to note that the `phase' boundary between 
Neel and spiral for spin-$1$ is closer to the the classical ($S \rightarrow 
\infty$) boundary $4J_2 = 1 - \delta^2$ than for spin-$1/2$. For instance, 
the cross-over from Neel to spiral occurs, for $\delta =0$, at $J_2 = 0.5$ 
for spin-$1/2$, at $0.39$ for spin-$1$, and at $0.25$ classically.

To summarize, we have studied a two-parameter `phase' diagram for 
the ground state of isotropic antiferromagnetic spin-$1/2$ and spin-$1$ 
chains using the NLSM approach, and have compared the results 
with those obtained numerically. We find that the spin-$1$ diagram is
considerably more complex than the corresponding spin-$1/2$ chain 
with surprising features like a `gapless' point inside the 
spiral `phase'; this point could be close to a critical point discussed 
earlier in the literature \cite{affl2,suth}. It would be interesting to 
establish this more definitively. 

Our results show that frustrated spin chains with small values of $S$ 
exhibit some features not anticipated from large $S$ field theories like the
NLSMs. The NLSMs also leave many questions unanswered. For instance, the
$O(3)$ NLSM which is applicable in the Neel phase does not tell us the
exponent of the gap which opens up as one moves away from $\theta =\pi$ (for
$g <g_c$) or as we go across $g=g_c$ (for $\theta =\pi$). To address these
questions, we have to use the more powerful technique of bosonization.

The NLSM approach can also be used to study spin chains in the presence
of a magnetic field. Consider adding a Zeeman term to the Hamiltonian in
(\ref{ham1}), i.e.,
\beq
H ~=~ \sum_i ~[ ~1 ~-~ (-1)^i ~\delta ~] ~ {\vec S}_i \cdot 
{\vec S}_{i+1} ~ + ~J_2 ~\sum_i ~{\vec S}_i \cdot {\vec S}_{i+2} ~-~ \sum_i ~
{\vec B} \cdot {\vec S}_i ~, 
\eeq
where $\vec B$ denotes the magnetic field.
In the region $1 - \delta^2 > 4J_2$, the classical ground state of this 
Hamiltonian is given by a coplanar configuration in which the spins 
${\vec S}_{2i}$ and ${\vec S}_{2i+1}$ lie at angles 
$\pm ~\alpha$ respectively with respect to the magnetic field, so that the
angle between the spins ${\vec S}_{2i}$ and ${\vec S}_{2i+1}$ is $2\alpha$.
Minimization of the energy fixes the angle $\alpha$ to be 
\beq
\alpha ~=~ \cos^{-1} \Bigl( ~\frac{|{\vec B}|}{4S} ~\Bigr) ~.
\label{defal}
\eeq
(We are assuming that $|{\vec B}| < 4S$, otherwise all the spins will align
with the magnetic field and $\alpha$ will be zero). We now define
\bea
{\vec \phi}_n ~&=&~ \frac{{\vec S}_{1n} ~-~ {\vec S}_{2n}}{2S \sin \alpha} ~, 
\nonu \\
{\vec l}_n ~&=&~ \frac{{\vec S}_{1n} ~+~ {\vec S}_{2n}}{2a} ~.
\label{phil2}
\eea
Note that the definition of $\vec \phi$ is slightly different from the one in 
(\ref{phil1}) in order to ensure that $\vec \phi$ is an unit vector. However,
$\vec l$ is orthogonal to and has the same commutation relations with $\vec 
\phi$ as before. We can now go to the continuum limit and derive the 
Hamiltonian
\beq
H ~=~ \int ~dx ~[~ \frac{cg^2}{2} ~({\vec l} + \frac{\theta}{4\pi} 
{\vec \phi}^{\prime})^2 ~ +~ \frac{c}{2g^2} ~{\vec \phi}^{\prime 2} -{\vec B}
\cdot {\vec l} ~]~, 
\eeq
where 
\bea
c ~&=&~ 2aS \sin \alpha ~{\sqrt {1 - 4J_2 - \delta^2}} ~, \nonu \\
g^2 ~&=&~ \frac{2}{S \sin \alpha {\sqrt {1 - 4 J_2 - \delta^2}}} ~, \nonu \\
{\rm and} \quad \theta ~&=&~ 2 \pi S \sin \alpha ~(1 - \delta) ~.
\eea
We can show that this follows from the Lagrangian density
\bea
{\cal L} ~&=&~ {1 \over {2 c g^2}} ~[~ {\vec \phi} \times \dot {\vec \phi} + 
{\vec B} - ({\vec B} \cdot {\vec \phi}) {\vec \phi} ~]^2 ~-~ {c \over {2 
g^2}} ~{\vec \phi}^{\prime 2} ~+~ {\theta \over {4 \pi}} ~{\vec \phi} 
\cdot {\vec \phi}^{\prime} \times {\dot {\vec \phi}} ~, \nonu \\
&=&~ {1 \over {2 c g^2}} ~[~ {\dot {\vec \phi}}^2 + 2 {\vec B} \times {\vec 
\phi} \cdot {\dot {\vec \phi}} + {\vec B}^2 - ({\vec B} \cdot {\vec \phi})^2 ~
] ~-~ {c \over {2 g^2}} ~{\vec \phi}^{\prime 2} ~+~ {\theta \over {4 \pi}} ~
{\vec \phi} \cdot {\vec \phi}^{\prime} \times {\dot {\vec \phi}} ~.
\label{lag2}
\eea
We see from this that
\beq
{\vec l} ~=~ {\vec \phi} \times {\vec \Pi} ~=~ \frac{1}{cg^2} ~[~{\vec \phi} 
\times \dot {\vec \phi} ~+~ {\vec B} ~-~ ({\vec B} \cdot {\vec \phi}) {\vec
\phi} ~] ~-~ \frac{\theta}{4\pi} ~{\vec \phi}^{\prime} ~.
\eeq
Since $cg^2 =4a$ and ${\vec B} \cdot {\vec \phi} =0$ in the classical ground 
state, we see that $\vec l$ is equal to ${\vec B}/(4a)$ plus small 
fluctuations; this agrees with its definition in (\ref{phil2}) and the 
classical configuration of the spins. One can now analyse the 
field theory governed by (\ref{lag2}) using the renormalization group and 
other methods. We refer the reader to \cite{norm} for further details. 

\section{Bosonization}

A very useful method for studying spin systems in one dimension is the
technique of bosonization. Before describing this method, let us briefly
present some background information. Further details can be found in Refs.
\cite{schu1,affl1,gogo,shan}. 

In one dimension, a great variety of interacting quantum systems (both
fermionic and bosonic) is described by the Tomonaga-Luttinger liquid (TLL) 
theory. Typically, a TLL describes quantum systems 
which are translation invariant and gapless, i.e.,
the excitation energy above the ground state is zero in the limit of the
system size $L \rightarrow \infty$. A TLL differs in three significant ways 
from the well-known Fermi liquid theory which describes many fermionic systems 
in two and three dimensions. First, {\it all} the low-energy excitations in a 
TLL have the character of sound modes which are bosonic and have a linear 
dispersion relation between the energy and the momentum (with the constant of 
proportionality being the sound velocity $v$). Even if the underlying theory 
is fermionic, the low-energy excitations are given by particle-hole pairs 
which are bosonic. The properties of a TLL are governed by two important 
parameters, namely, an interaction parameter $K$ (noninteracting systems 
have $K=1$) and the velocity $v$. Secondly, the one-particle 
momentum distribution function $n (k)$ for fermions, which is obtained 
by Fourier transforming the fermion Green's function 
\beq
G (x, t) ~=~ < 0 | T \psi (x, t) \psi^{\dagger} (0, 0) |
0> 
\eeq
and computing the residue of its pole in the complex $\omega$ plane as a 
function of $k$, has no discontinuity at the Fermi surface $k=k_F$ for a
TLL. Instead, it has a cusp there of the form
\beq
n (k) ~=~ n(k_F) ~+~ {\rm const.} ~{\rm sign} (k-k_F) ~|k-k_F|^{(1-K)^2/2K} ~.
\eeq
On the other hand, in a Fermi liquid, $n(k)$ has a finite discontinuity at the
Fermi surface; see Fig. 6. Finally, correlation functions in a TLL typically 
decay at large distances as power-laws which depend on $K$, unlike the 
correlation functions of a Fermi liquid where the power-laws are universal.
 
\begin{figure}[htb]
\begin{center}
\epsfig{figure=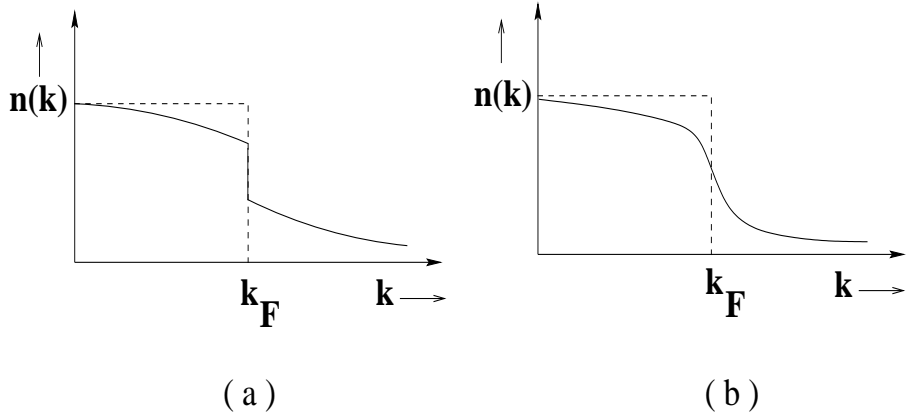,height=5.5cm,width=12cm}
\end{center}
\caption{One-particle momentum distribution function for (a) an interacting 
Fermi liquid, and (b) a Tomonaga-Luttinger liquid.}
\label{fig6}
\end{figure}

Let us be more specific about the nature of the low-energy excitations in a 
one-dimensional system of interacting fermions. Assume that we have a system 
of length $L$ with periodic boundary conditions; the translation invariance 
and the finite length make the one-particle momenta discrete. Suppose that the 
system has $N_0$ particles with a ground state energy $E_0 (N_0)$ and a ground 
state momentum $P_0 =0$. We will be interested
in the thermodynamic limit $N_0 , L \rightarrow \infty$ keeping the particle 
density $\rho_0 = N_0 /L$ fixed. If we could switch off the interactions,
the fermions would have two Fermi points, at $k = \pm k_F$ respectively, with
all states with momenta lying between the two points being occupied.
(See Fig. 7 for a typical picture of the momentum states of a lattice model
without interactions). Even in the presence of interactions, it turns out that 
the low-lying excitations consist of two pieces \cite{hald2},

\noindent (i) a set of bosonic excitations each of which can have either 
positive momentum $q$ or negative momentum $-q$ with an energy $\epsilon_q 
= v q$, where $0 < q << k_F$ and $v$ is the sound velocity, and

\noindent (ii) a certain number of particles $N_R$ and $N_L$ added to the
right and left Fermi points respectively, where $N_R , N_L << N_0$. Note
that $N_R$ and $N_L$ can be positive, negative or zero. 

The quasiparticle excitations in (i) have an infinite number of degrees
of freedom (in the thermodynamic limit), and they determine properties such
as specific heat and susceptibility to various perturbations. The particle
excitations in (ii) only have two degrees of freedom and therefore play no
role in the thermodynamic properties.
The Hamiltonian and momentum operators for a 
one-dimensional system (which may have interactions) have the general form 
\bea
H ~= && E_0 (N_0) ~+~ \sum_{q>0} ~vq~ {[} ~{\tilde b}_{R,q}^\dagger 
{\tilde b}_{R,q} ~+~ 
{\tilde b}_{L,q}^\dagger {\tilde b}_{L,q} ~{]} ~\nonu \\
&& +~ \mu (N_R + N_L ) ~+~ \frac{\pi v}{2LK} (N_R + N_L )^2 ~+~ \frac{\pi 
vK}{2L} ~(N_R - N_L )^2 ~, \nonu \\
P ~= && \sum_{q>0} ~q~ {[} ~{\tilde b}_{R,q}^\dagger {\tilde b}_{R,q} ~-~ 
{\tilde b}_{L,q}^\dagger {\tilde b}_{L,q} ~{]} ~+~ [~ k_F ~+~ 
\frac{\pi}{L} (N_R + N_L ) ~]~ (N_R - N_L ) ~,
\label{xham1}
\eea
where $q$ is the momentum of the low-energy bosonic excitations created and 
annihilated by ${\tilde b}_q^\dagger$ and ${\tilde b}_q$,
$K$ is a positive dimensionless number, and $\mu$ is the chemical potential
of the system. We will see later that $v$ and $K$ are the two important 
parameters which determine all the low-energy properties of a system. Their 
values generally depend on both the strength of the interactions
and the density. If the fermions are noninteracting, we have
\beq
v ~=~ v_F \quad {\rm and} \quad K ~=~ 1 ~.
\label{vk}
\eeq
Note that one can numerically find the values of $v$ and $K$ by varying
$N_R$ and $N_L$ and studying the $1/L$ dependence of energy and momentum of 
finite size systems.

The technique of bosonization (combined with conformal field theory) is very 
useful for analytically studying a TLL \cite{schu1,affl1,gogo,shan}. This 
technique consists of mapping bosonic operators into fermionic ones, and then
using whichever set of operators is easier to compute with. 

To begin, let us consider a fermion with both right- and left-moving 
components. We introduce a chirality label $\nu$, such that $\nu = R$ and
$L$ refer to right- and left-moving particles respectively. Sometimes we will
use the numerical values $\nu = 1$ and $-1$ for $R$ and $L$; this will be 
clear from the context. Then the second quantized Fermi fields are given by
\bea
\psi_\nu (x) ~&=&~ \frac{1}{\sqrt L} ~\sum_{k=-\infty}^{\infty} ~
c_{\nu,k} ~e^{i\nu kx} ~, \nonu \\
k ~&=&~ \frac{2\pi}{L} ~n_k ~,
\label{psi2}
\eea
where $n_k = 0, \pm 1, \pm 2, ...$, and
\bea
\{ c_{\nu ,k} , c_{\nu^\prime , k^\prime} \} ~&=&~ 0 ~, \nonu \\
\{ c_{\nu ,k} , c_{\nu^\prime , k^\prime}^\dagger \} ~&=&~ 
\delta_{\nu \nu^\prime} ~\delta_{k k^\prime} ~. 
\eea

Next we define bosonic operators
\bea
b_{\nu ,q}^\dagger ~&=&~ \frac{1}{\sqrt n_q} ~\sum_{k=-\infty}^{\infty} ~
c_{\nu ,k+q}^\dagger c_{\nu ,k} ~, \nonu \\
b_{\nu ,q} ~&=&~ \frac{1}{\sqrt n_q} ~\sum_{k=-\infty}^{\infty} ~
c_{\nu ,k-q}^\dagger c_{\nu ,k} ~.
\eea
Note that $b_{R,q}^\dagger$ and $b_{L,q}^\dagger$ create excitations with
momenta $q$ and $-q$ respectively, where the label $q$ is
always taken to be positive. We can show that
\beq
{[} b_{\nu ,q} , b_{\nu^\prime , q^\prime} {]} ~=~ 0 ~, \quad {\rm 
and} \quad {[} b_{\nu ,q} , b_{\nu^\prime , q^\prime}^\dagger {]} ~=~ 
\delta_{\nu \nu^\prime} ~\delta_{q q^\prime} ~.
\eeq
The vacuum state of the system is defined to be the state $|0>$ which is 
annihilated by the operators $c_{\nu,k}$ for $k \ge 0$ and 
$c_{\nu,k}^{\dagger}$ for $k < 0$, and therefore by $b_{\nu,q}$ for all $q$. 

Let us define the chiral bosonic fields
\beq
\phi_\nu (x) ~=~ \frac{i\nu}{2 {\sqrt \pi}} ~\sum_{q>0} ~\frac{1}{\sqrt 
n_q} ~[~ b_{\nu ,q} ~e^{i\nu qx - \alpha q/2} ~-~ b_{\nu ,q}^\dagger ~e^{-i
\nu qx - \alpha q/2} ~] ~-~ \frac{\sqrt \pi x}{L} ~{\hat N}_\nu ~,
\label{chibos}
\eeq
where the length parameter $\alpha$ is a cut-off which is required to ensure 
that the contribution from high-momentum modes do not produce divergences 
when computing correlation functions. The fields in (\ref{chibos}) satisfy
\beq
{[} \phi_\nu (x) , \phi_{\nu^\prime} (x) {]} ~=~ - \frac{i\nu}{4} ~
\delta_{\nu \nu^\prime} ~{\rm sign} ~(x-x^\prime ) 
\eeq
in the limit $\alpha \rightarrow 0$. It is useful to define two fields 
dual to each other
\bea
\phi (x) ~&=&~ \phi_R (x) ~+~ \phi_L (x) ~, \nonu \\
\theta (x) ~&=&~ -~ \phi_R (x) ~+~ \phi_L (x) ~.
\eea
Then $[ \phi (x), \phi (x^\prime)] = [ \theta (x), \theta (x^\prime)] 
= 0$, while
\beq
{[} \phi (x) , \theta (x^\prime) {]} ~=~ \frac{i}{2} ~{\rm sign} ~(x-
x^\prime) ~.
\eeq

Now it can be shown that the fermionic and bosonic operators discussed above
are related to each other as
\bea
\psi_R ~&=&~ \frac{1}{\sqrt {2\pi \alpha}} ~\eta_R ~e^{-i 2 {\sqrt \pi} 
\phi_R} ~, \nonu \\
\psi_L ~&=&~ \frac{1}{\sqrt {2\pi \alpha}} ~\eta_L ~e^{i 2 {\sqrt \pi} 
\phi_L} ~,
\label{bos}
\eea
in the sense that they produce the same state when they act on the vacuum 
state $|0>$, and they have the same correlation functions. The 
unitary operators $\eta_R$ and $\eta_L$ are called Klein factors, and they are 
essential to ensure that the fermionic fields given in Eq. (\ref{bos}) 
anticommute at two different spatial points $x$ and $y$. 

The densities of the right- and left-moving fermions are given by $\rho_R = 
\psi_R^\dagger \psi_R$ and $\rho_L = \psi_L^\dagger \psi_L$.
The total fermionic density and current are given by
\bea
\rho ~-~ \rho_0 ~&=&~ \rho_R + \rho_L ~=~ - ~\frac{1}{\sqrt \pi} ~
\frac{\partial \phi}{\partial x} ~, \nonu \\
j ~&=&~ v_F ~(\rho_R - \rho_L ) ~=~ \frac{v_F}{\sqrt \pi} ~\frac{\partial 
\theta}{\partial x} ~,
\eea
where $\rho_0$ is the background density (fluctuations
around this density are described by the fields $\psi$ or $\phi$), and
the velocity $v_F$ will be introduced below.

Let us now introduce a Hamiltonian.
We assume a linear dispersion relation $\epsilon_{\nu ,k} = 
v_F k$ for the fermions. The noninteracting Hamiltonian then takes the form
\bea
H_0 &=& v_F \sum_{k=-\infty}^\infty ~k ~{[} ~c_{R,k}^\dagger c_{R,k} ~+~ 
c_{L,k}^\dagger c_{L,k} ~{]} ~+~ \frac{\pi v_F}{L} ({\hat N}_R^2 + {\hat 
N}_L^2 ) \nonu \\
&=& -v_F \int_0^L dx ~[ \psi_R^\dagger (x) i \partial_x \psi_R (x) ~-~
\psi_L^\dagger (x) i \partial_x \psi_L (x) ] ~+~ \frac{\pi v_F}{L} 
({\hat N}_R^2 + {\hat N}_L^2 )
\eea
in the fermionic language, and
\bea
H_0 ~&=&~ v_F ~\sum_{q>0} ~q ~(~ b_{R,q}^\dagger b_{R,q} ~+~ 
b_{L,q}^\dagger b_{L,q} ~)~ +~ \frac{\pi v_F}{L} ({\hat N}_R^2 + 
{\hat N}_L^2 ) \nonu \\
&=&~ v_F ~\int_0^L ~dx ~[ ~(\partial_x \phi_R )^2 ~+~ (\partial_x \phi_L 
)^2 ~] \nonu \\
&=&~ \frac{v_F}{2} ~\int_0^L ~dx ~[ ~(\partial_x \phi )^2 ~+~ (\partial_x 
\theta )^2 ~]
\eea
in the bosonic language. 

We now study the effects of four-fermi interactions. Let us 
consider an interaction of the form
\beq
V ~=~ \frac{1}{2} ~\int_0^L ~dx ~[~ 2g_2 ~\rho_R (x) \rho_L (x) ~+~g_4 ~
(~ \rho_R^2 (x) ~+~ \rho_L^2 (x) ~) ~] ~.
\eeq
Physically, we may expect an interaction such as $g \rho^2 /2$, so that
$g_2 = g_4 =g$. However, it is instructive to allow $g_2$ to differ from 
$g_4$ to see what happens. Also, we will 
not assume anything about the signs of $g_2$ and $g_4$. In the fermionic
language, the interaction takes the form
\bea
V= \frac{1}{2L} \sum_{k_1 , k_2 ,k_3 =-\infty}^{\infty} &[& 2 g_2 
c_{R,k_1 + k_3}^\dagger c_{R,k_1} c_{L,k_2 +k_3}^\dagger c_{L,k_2} \nonu \\
&& ~+ g_4 ( c_{R,k_1 + k_3}^\dagger c_{R,k_1} c_{R,k_2 -
k_3}^\dagger c_{R, k_2} + c_{L,k_1 + k_3}^\dagger c_{L,k_1} c_{L,k_2 -
k_3}^\dagger c_{L,k_2} )] ~. \nonu \\
&&
\label{int}
\eea
{} From this expression we see that $g_2$ corresponds to a two-particle 
scattering involving both chiralities; in this model, we can call it 
either forward scattering or backward scattering since there is no 
way to distinguish between the two processes
in the absence of some other quantum number such as spin. 
The $g_4$ term corresponds to a scattering between two fermions with the
same chirality, and therefore describes a forward scattering process. 

The quartic interaction in Eq. (\ref{int}) seems very difficult to analyze. 
However we will now see that it is easily solvable in the bosonic 
language; indeed this is one of the main motivations behind bosonization. 
The bosonic expression for the total Hamiltonian $H = H_0 +V$ is found to be 
\bea
H &=& \sum_{q>0} ~q {[} v_F (b_{R,q}^\dagger b_{R,q} + 
b_{L,q}^\dagger b_{L,q} )+ \frac{g_2}{2\pi} (b_{R,q}^\dagger 
b_{L,q}^\dagger + b_{R,q} b_{L,q} ) + \frac{g_4}{2\pi} (b_{R,q}^\dagger 
b_{R,q} + b_{L,q}^\dagger b_{L,q} ) {]} \nonu \\
&& ~~+ \frac{\pi v_F}{L} (~ {\hat N}_R^2 + {\hat N}_L^2 ~) 
+~ \frac{g_2}{L} ~{\hat N}_R {\hat N}_L ~+~ \frac{g_4}{2L} (~
{\hat N}_R^2 + {\hat N}_L^2 ~) .
\eea
The $g_4$ term only renormalizes the velocity. The $g_2$ term can then be
rediagonalized by a Bogoliubov transformation. We first define two
parameters
\bea
v ~&=&~ \Bigl[ ~(~ v_F + \frac{g_4}{2\pi} - \frac{g_2}{2\pi} ~)~
(~ v_F + \frac{g_4}{2\pi} + \frac{g_2}{2\pi} ~) ~\Bigr]^{1/2} ~, \nonu \\
K ~&=&~ \Bigl[ ~(~ v_F + \frac{g_4}{2\pi} - \frac{g_2}{2\pi}~) ~/~ (~v_F + 
\frac{g_4}{2\pi} + \frac{g_2}{2\pi} ~) ~\Bigr]^{1/2} ~.
\eea
Note that $K < 1$ if $g_2$ is positive (repulsive interaction), and $> 1$ 
if $g_2$ is negative (attractive interaction). [If $g_2$ is so large that
$v_F + g_4/(2\pi) - g_2/(2\pi) < 0$, then our analysis breaks down. The 
system does not remain a Luttinger liquid in that case, and is likely to go 
into a different phase such as a state with charge density order]. 
The Bogoliubov transformation then takes the form
\bea
{\tilde b}_{R,q} ~&=&~ \frac{b_{R,q} ~+~ \gamma ~
b_{L,q}^{\dagger}}{\sqrt {1 - \gamma^2}} ~, \nonu \\
{\tilde b}_{L,q} ~&=&~ \frac{b_{L,q} ~+~ \gamma ~
b_{R,q}^{\dagger}}{\sqrt {1 - \gamma^2}} ~, \nonu \\
{\rm where} \quad \gamma ~&=&~ \frac{1-K}{1+K} ~,
\eea
for each value of the momentum $q$. The Hamiltonian is then given by
the quadratic expression
\bea
H ~=~ & & \sum_{q>0} ~v q ~{[}~ {\tilde b}_{R,q}^{\dagger} {\tilde 
b}_{R,q} ~+~ {\tilde b}_{L,q}^{\dagger} {\tilde b}_{L,q} ~{]} \nonu \\
& & +~ \frac{\pi v}{2L} ~[~ \frac{1}{K} ~({\hat N}_R + {\hat N}_L )^2 ~+~ K~
({\hat N}_R - {\hat N}_L )^2 ~]~ .
\label{xham5}
\eea
Equivalently,
\beq
H ~=~ \frac{1}{2} ~\int_0^L ~dx ~[~ vK \Pi^2 ~+~ \frac{v}{K} ~(\partial_x
\phi)^2 ~] ~.
\label{xham6}
\eeq
The old and new fields are related as
\bea
\phi_R ~&=&~ \frac{(1+K) ~\tphi_R ~-~ (1-K) ~\tphi_L}{2 {\sqrt K}} ~, 
\nonu \\
\phi_L ~&=&~ \frac{(1+K) ~\tphi_L ~-~ (1-K) ~\tphi_R}{2 {\sqrt K}} ~, 
\nonu \\
\phi ~&=&~ {\sqrt K} ~\tphi \quad {\rm and} \quad \theta ~=~ 
\frac{1}{\sqrt K} ~{\tilde \theta} ~.
\label{oldnew}
\eea

Note the important fact that the vacuum changes as a result of the 
interaction; the new vacuum $\vert {\tilde 0} \rangle$ is the state 
annihilated by the 
operators ${\tilde b}_{\nu ,q}$. Since the various correlation functions
must be calculated in this new vacuum, they will depend on the interaction
through the parameters $v$ and $K$. In particular, we will see below that 
the power-laws of the correlation functions are governed by $K$.

Given the various Hamiltonians, it is easy to guess the forms of the 
corresponding Lagrangians. For the noninteracting theory ($g_2 = g_4 =0$), 
the Lagrangian density describes a massless Dirac fermion,
\beq
{\cal L} ~=~ i \psi_R^\dagger (\partial_t + v_F \partial_x ) \psi_R ~ +~ i 
\psi_L^\dagger (\partial_t - v_F \partial_x ) \psi_L 
\eeq
in the fermionic language, and a massless real scalar field,
\beq
{\cal L} ~=~ \frac{1}{2v_F} ~(\partial_t \phi )^2 ~ -~ \frac{v_F}{2} ~( 
\partial_x \phi )^2 
\eeq
in the bosonic language. For the interacting theory in Eq. (\ref{xham6}), we 
find from Eq. (\ref{oldnew}) that
\beq
{\cal L} ~=~ \frac{1}{2vK} ~(\partial_t \phi )^2 ~-~ \frac{v}{2K} ~( 
\partial_x \phi )^2 ~=~ \frac{1}{2v} ~(\partial_t \tphi )^2 ~-~ 
\frac{v}{2} ~( \partial_x \tphi )^2 ~.
\eeq

Although the dispersion relation is generally not linear for all the modes of
a realistic system, it often happens that the low-energy and long-wavelength 
modes (and therefore the low-temperature properties) can be 
described by a TLL. For a fermionic system in one dimension, these modes are
usually the ones lying close to the two Fermi points with momenta $\pm k_F$ 
respectively; see Fig. 7. Although the fermionic field $\psi$ generally has 
components with all possible momenta, one can define right- and left-moving
fields $\psi_R$ and $\psi_L$ which vary slowly on the scale length $a$,
\beq
\psi (x,t) ~=~ \psi_R (x,t) ~e^{ik_F x} ~+~ \psi_L (x,t) ~e^{-ik_F x} ~.
\eeq
Quantities such as the density generally contain terms which vary
slowly as well as terms varying rapidly on the scale of $a$,
\bea
\rho - \rho_0 ~&=&~ \psi^\dagger \psi ~=~ \psi_R^\dagger \psi_R ~+~ 
\psi_L^\dagger \psi_L ~+~ e^{-i2k_F x} ~\psi_R^\dagger \psi_L ~+~ 
e^{i2k_F x} ~ \psi_L^\dagger \psi_R \nonu \\
&=& ~- ~\frac{1}{\sqrt \pi} ~\frac{\partial \phi}{\partial x} ~+~\frac{1}{2\pi
\alpha} ~{[} \eta_R^\dagger \eta_L e^{i(2 {\sqrt \pi} \phi - 2k_F x)} ~+~
\eta_L^\dagger \eta_R e^{-i(2 {\sqrt \pi} \phi - 2k_F x)} ~{]} .
\label{den}
\eea

\begin{figure}[htb]
\begin{center}
\epsfig{figure=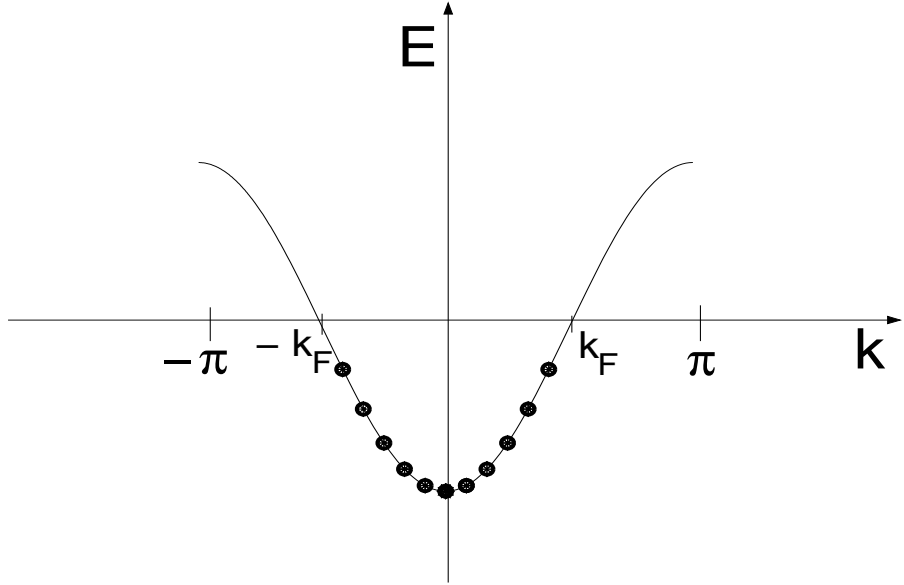,height=8cm,width=12cm}
\end{center}
\caption{Picture of the ground state of a one-dimensional system of
noninteracting fermions on a lattice. Filled circles denote occupied states
lying below the Fermi energy $E_F =0$.}
\label{fig7}
\end{figure}

One can now compute various correlation functions in the bosonic language. 
Consider an operator of the exponential form
\beq
O_{m,n} ~=~ e^{i2 {\sqrt \pi} (m \phi + n \theta)} ~.
\eeq
(Such an operator can arise from a product of several $\psi$'s and 
$\psi^\dagger$'s if we ignore the Klein factors; then Eq. (\ref{bos}) 
implies that $m \pm n$ must take integer values). We then find the following 
result for the two-point correlation function at space-time separations
which are much larger than the microscopic lattice spacing $a$,
\bea
&& \langle {\tilde 0} \vert ~T O_{m,n} (x,t) O_{m^\prime , 
n^\prime }^\dagger (0,0) ~\vert {\tilde 0} \rangle \nonu \\
&& \sim \delta_{mm^\prime } \delta_{nn^\prime } \frac{\alpha^{2(m^2 K + 
n^2 /K)}}{(vt-x-i\alpha {\rm sign} ( t))^{(m {\sqrt K} - n/ {\sqrt K})^2 } 
(vt+x-i\alpha {\rm sign} (t))^{(m{\sqrt K} + n/{\sqrt K})^2} } ~. \nonu \\
&&
\label{corr}
\eea
Note that the correlation function decays as a power-law, and the power 
depends on the interaction parameter $K$. In the language of the 
renormalization group, the scaling dimension of $O_{m,n}$ is given by
\beq
d_O ~=~ m^2 K ~+~ \frac{n^2}{K} ~.
\eeq

We can now discuss a spin chain from the point of view of bosonization.
To be specific, let us consider a spin-$1/2$ chain described by the 
anisotropic Hamiltonian 
\beq
H ~=~ \sum_{i=1}^N ~[~ \frac{J}{2} ~(S_i^+ S_{i+1}^- + S_i^- S_{i+1}^+) ~+~
J \Delta S_i^z S_{i+1}^z ~-~ h S_i^z ~] ~,
\label{ham3}
\eeq
where the interactions are only between nearest neighbor spins, and
$J > 0$. $S_i^+= S_i^x+iS_i^y$ and $S_i^-=S_i^x-iS_i^y$ are the spin 
raising and lowering operators, and $h$ denotes a magnetic field. 
Note that the model has a $U(1)$ invariance, namely, rotations 
about the $S^z$ axis. When $\Delta =1$ and $h=0$, the $U(1)$ invariance 
is enhanced to an $SU(2)$ invariance, because at this point the model can 
be written simply as $H = J \sum_i {\vec S_i} \cdot {\vec S_{i+1}}$. 

Eq. (\ref{ham3}) is the well-studied $XXZ$ spin-$1/2$ chain in a 
longitudinal magnetic field. It can be exactly solved using the Bethe ansatz, 
and a lot of information can then be obtained using conformal field theory 
\cite{cabr,hald2}. The following results are relevant for us. The model is 
gapless for a certain range of values of $\Delta$ and $h/J$.
For instance, this is true if $-1 < \Delta \le 1$ and $h=0$; then the two-spin 
equal-time correlations have oscillatory pieces which decay asymptotically as
\bea
\langle S_0^+ S_n^- \rangle ~& \sim &~ \frac{(-1)^n}{\vert n \vert^\eta} ~,
\nonu \\
\langle S_0^z S_n^z \rangle ~& \sim &~ \frac{(-1)^n}{\vert n \vert^{1/\eta}}~,
\nonu \\
{\rm where} \quad \eta ~&=&~ \frac{1}{2} ~+~ \frac{1}{\pi} ~\sin^{-1} ~ 
(\Delta ) ~.
\label{corrxxz}
\eea
For $\Delta > 1$ and $h=0$, the system is gapped; there are two
degenerate ground states which have a period of two sites consistent with
the condition (\ref{quant}). Thus the invariance of the Hamiltonian under a 
translation by one site is spontaneously broken in the ground states. 
This is particularly obvious for $\Delta \rightarrow \infty$ 
where the two ground states are $+-+- \cdots$ and $-+-+ \cdots$. The
two-spin correlations decay exponentially for $\Delta > 1$ and $h=0$. 
Finally, the system is gapped for $h/J > 1+ \Delta$ with all sites having 
$S_z = 1/2$ in the ground state, and for $h/J < -1- \Delta$ with all sites 
having $S_z = -1/2$. 

However, it is not easy to compute explicit correlation functions using the 
Bethe ansatz. We will therefore use bosonization to study the model in 
(\ref{ham3}).

We first use the Jordan-Wigner transformation to map the spin model to a model 
of spinless fermions. We map an $\uparrow$ spin or a $\downarrow$ spin at 
any site to the
presence or absence of a fermion at that site. We introduce a fermion 
annihilation operator $\psi_i$ at each site, and write the spin at the site as
\bea
S_i^z &=& \psi_i^{\dagger} \psi_i -1/2 = n_i-1/2 \nonu \\
S_i^- &=& (-1)^i ~\psi_i e^{i\pi \sum_j n_j} ~,
\label{jorwig}
\eea 
where the sum runs from one boundary of the chain up to the $(i-1)^{\rm th}$ 
site (we assume an open boundary condition here for convenience), $n_i =0$ or 
$1$ is the fermion occupation number at site $i$, and the expression for 
$S_i^+$ is obtained by taking the hermitian conjugate of $S_i^-$, 
The string factor in the definition of $S_i^-$ is added in order to 
ensure the correct statistics for different sites; the fermion operators at 
different sites anticommute, whereas the spin operators commute. 

We now find that 
\beq
H ~=~ -~ ~\sum_i ~[~\frac{J}{2} ~(\psi_i^{\dagger}\psi_{i+1} +h.c.) ~+~
J \Delta ~(n_i-1/2)(n_{i+1}-1/2) ~-~ h ~(n_i -1/2) ~]~.
\eeq
We see that the spin-flip operators $S_i^{\pm}$ lead to hopping terms in 
the fermion Hamiltonian, whereas the $S_i^z S_{i+1}^z$ 
term leads to an interaction between fermions on adjacent sites. 

Let us first consider the noninteracting case given by $\Delta =0$.
By Fourier transforming the fermions, $\psi_k = \sum_j \psi_j
e^{-ikja}/\sqrt{N}$, where $a$ is the lattice
spacing and the momentum $k$ lies in the first Brillouin 
zone $-\pi /a < k \le \pi /a$, we find that the Hamiltonian is given by 
\beq
H ~=~ \sum_k ~\omega_k ~\psi_k^{\dagger}\psi_k ~,
\eeq 
where 
\beq
\omega_k ~=~ -J ~\cos (ka) ~-~ h~.
\eeq
The noninteracting ground state is the one in which all the single-particle
states with $\omega_k <0$ are occupied, and all the states with 
$\omega_k >0$ are empty. If we set the magnetic field $h=0$,
the magnetization per site $m \equiv \sum_i S_i^z /N$ will be zero in the
ground state; equivalently, in the fermionic 
language, the ground state is precisely half-filled. Thus, for $m=0$, 
the Fermi points ($\omega_k =0$) lie at $ka = \pm \pi/2 \equiv k_F a$.
Let us now add the magnetic field term. In the fermionic language, this is 
equivalent to adding a chemical potential term (which couples to $n_i$ or 
$S_i^z$). In that case, the ground state no longer has $m=0$ and the fermion 
model is no longer half-filled. The Fermi points are then given 
by $\pm k_F$, where 
\beq
k_F a ~=~ \pi ~( m ~+~ \frac{1}{2} ) ~.
\label{kf}
\eeq
It turns out that this relation between $k_F$ (which governs the oscillations 
in the correlation functions as discussed below) and the magnetization $m$ 
continues to hold even if we turn on the interaction $J\Delta$, although the 
simple picture of the ground state (with states filled below some energy and 
empty above some energy) no longer holds in that case.

In the linearized approximation, the modes near the two Fermi points have 
the velocities $\partial \omega_k / \partial k = \pm v$, where
$v$ is some function of $J$, $\Delta$ and $h$. Next, we introduce the slowly 
varying fermionic fields $\psi_R$ and $\psi_L$ as indicated above; these are 
functions of a coordinate $x$ which must be an integer multiple of $a$. Now 
we bosonize these fields. The spin fields can be written in terms 
of either the fermionic or the bosonic fields. For instance, $S^z$ is
given by the fermion density as in Eq. (\ref{jorwig}) which then has a 
bosonized form given in Eq. (\ref{den}). Similarly, 
\bea
S^+(x,t) ~&=&~ (-1)^{x/a} ~[ \psi_R^\dagger (x,t) e^{-ik_Fx/a} + 
\psi_L^\dagger (x,t) e^{ik_Fx/a} ] ~\times \nonu \\
&& ~~~~~~~~~~~~ [e^{i\pi \int_{-\infty}^{x} dx' (\psi^{\dagger}(x',t) 
\psi(x',t) + 1/2a)} + h.c. ]~,
\eea
where $(-1)^{x/a} = \pm 1$ since $x/a$ is an integer. This can now be written
entirely in the bosonic language; the term in the exponential is given by
\beq
\int_{-\infty}^x dx' \psi^{\dagger}(x',t)\psi(x',t) = -~{1 \over \sqrt{\pi}}
\int_{-\infty}^x dx' \partial_{x'} \phi ~=~ -~{1 \over \sqrt{\pi}} ~[\phi_R
(x,t) +\phi_L(x,t)]~,
\eeq
where we have ignored the contribution from the lower limit at $x' = -\infty$.

We can now use this bosonic expressions to compute the various two-spin
correlation functions $G^{ab} (x,t) \equiv <0| T S^a (x,t) S^b (0,0) | 0 >$. 
We find that
\bea
G^{zz} (x,t) & = & m^2 ~+~ c_1 \Bigl[ {1 \over (x+vt)^2} + {1 \over 
(x-vt)^2} \Bigr] + c_2 \frac{\cos (2k_F x)}{(x^2 - v^2 t^2)^K} ~, \nonu \\
G^{+-}(x,t) + G^{-+}(x,t) & = & c_3 \frac{(-1)^{x/a}}{(x^2 - v^2 
t^2)^{1/4K}} \nonu \\
&& + c_4 \frac{(-1)^{x/a}~\cos (2k_F x)}{(x^2 -v^2 t^2)^{({\frac{1}{2{\sqrt 
K}} - {\sqrt K}})^2}} \Bigl[ {1 \over (x-vt)^2} + {1 \over (x+vt)^2} \Bigr] ~,
\nonu \\
&&
\eea 
where $c_1, ..., c_4$ are some constants. The Luttinger parameters $K$ and 
$v$ are functions of $\Delta$ and $h/J$ (or $m$). [The exact dependence can 
be found from the web site given in Ref. \cite{cabr}; this contains a 
calculator which finds the values of $R=1/{\sqrt {4\pi K}}$ and $h/J$ if one 
inputs the values of $M=2m$ and $\Delta$]. For $h=0$, $K$ is given by the 
analytical expression 
\beq
{1 \over K} ~=~ 1 ~+~ {2 \over \pi} \sin^{-1} (\Delta ) ~.
\eeq
Note that at the $SU(2)$ invariant point $\Delta =1$ and $h=0$, we have 
$K=1/2$, and the two correlations $G^{zz}$ and $G^{+-}$ have the same forms.

In addition to providing a convenient way of computing correlation functions, 
bosonization also allows us to study the effects of small perturbations which 
may take the system away from a TLL. For instance, a physically important 
perturbation is a dimerizing term
\beq
V ~=~ \delta \sum_i ~(-1)^{i} ~[~ \frac{J}{2} ~(S_i^+ S_{i+1}^- + S_i^- 
S_{i+1}^+) ~+~ J \Delta S_i^z S_{i+1}^z ~] ~,
\eeq
where $\delta$ is the strength of the perturbation. Upon bosonizing, we find
that the scaling dimension of this term is $K$. Hence it is relevant if
$K < 2$; in that case, it produces an energy gap in the system which scales
with $\delta$ as
\beq
\Delta E ~\sim ~\delta^{1/(2-K)} ~.
\eeq
For the isotropic case $\Delta =1$, we have $K=1/2$ and the gap scales as 
$\Delta E \sim \delta^{2/3}$. [This is the exponent of the gap which appears
as we vary $\delta$ to move away from the gapless line $(0\le J_2 \le J_{2c}, 
\delta =0)$ for spin-$1/2$ in Fig. 2 or the line A for spin-$1$ in Fig. 3]. 
This phenomenon occurs in spin-Peierls systems such as $CuGeO_3$; below a 
transition temperature $T_{sp}$, they go into a dimerized phase which has a 
gap.

Another interesting perturbation occurs when the frustration parameter $J_2$ 
crosses the critical value $J_{2c} =0.241$ for $\delta =0$ in the spin-$1/2$ 
chain; see Fig. 2. This turns out to be a marginal perturbation, and it 
produces a gap which has an essential singularity of the form $\Delta E \sim 
\exp [- {\rm const.}/(J_2 - J_{2c})]$ \cite{okam}. Because of this form, it
is very hard to numerically measure the gap if $J_2$ is close to $J_{2c}$.

Finally, when two isotropic spin-$1/2$ chains (with the spin variables
in the two chains being denoted by ${\vec S}^{(1)}_n$ and ${\vec S}^{(2)}_n$) 
are coupled together with a weak interchain coupling 
\beq
V ~=~ \jp ~\sum_n ~{\vec S}^{(1)}_n \cdot {\vec S}^{(2)}_n ~, 
\eeq
we find that the perturbation ${\vec S}^{(1)}
\cdot {\vec S}^{(2)}$ has the scaling dimension $1$. Hence this perturbation
is relevant, and it produces an energy gap which scales as $\Delta E \sim
\jp$. This has been confirmed by numerical calculations \cite{pati}.

\section{Low-energy Effective Hamiltonian approach}

As mentioned in Sec. 1, a quantum spin system can sometimes
exhibit magnetization plateaus. For a Hamiltonian which is invariant under 
translation by one unit cell, the value of the magnetization per unit cell is 
quantized to be a rational number at each plateau. 
The necessary (but not sufficient) condition for the magnetization
quantization is given as follows \cite{oshi}. Let us assume that the magnetic 
field points along the $\hat z$ axis, the total Hamiltonian $H$ is invariant 
under spin rotations about that axis, and the maximum possible spin in each 
unit cell of the Hamiltonian is given by $S$. Consider a state $\psi$ such
that the expectation value of $S_z$ per unit cell is equal to $m_s$ in that 
state, and $\psi$ has a period $n$, i.e., it is invariant only under 
translation by a number of unit cells equal to $n$ or a multiple of $n$. (It 
is clear that if $n \ge 2$, then there must be $n$ such states with the same
energy, since $H$ is invariant under a translation by one unit cell). Then
the quantization condition says that a magnetic plateau is possible, i.e., 
there is a range of values of the external field for
which $\psi$ is the ground state and is separated by a finite gap from states
with slightly higher or lower values of total $S_z$, only if 
\beq
n ~(~ S ~-~ m_s ~) ~=~ {\rm an ~~ integer} .
\label{quant}
\eeq
Note that the saturated state in which all spins point along the magnetic 
field trivially satisfies (\ref{quant}) since it has $m_s =S$ (or $-S$) 
and $n=1$.

In this section, we study the magnetization as a function of the applied field 
for a two- and three-chain ladder using a perturbatively derived low-energy 
effective Hamiltonian (LEH) \cite{tots1,mila}. In both cases, the first-order 
LEH will turn out to be the model described in Eq. (\ref{ham3}). As we pointed 
out earlier, a lot is known about this model \cite{cabr,hald2}. 
In particular, we will see that the exponent $\eta$ for the correlation power 
laws can be read off from the expression for the first-order LEH. 

We consider a three-chain spin-$1/2$ ladder governed by the Hamiltonian 
\beq
H ~=~ \jp ~\sum_a ~\sum_n ~ {\vec S}_{a,n} \cdot {\vec S}_{a+1,n} ~+~ J ~
\sum_{a=1}^3 ~\sum_n ~{\vec S}_{a,n} \cdot {\vec S}_{a,n+1} ~
-~ h ~\sum_{a=1}^3 ~\sum_n ~S_{a,n}^z ~,
\label{ham4}
\eeq
where $a$ denotes the chain index, $n$ denotes the rung index, $h$ denotes 
the magnetic field, and $J, \jp > 0$; see Fig. 8. 
We may choose $h \ge 0$ since the region $h < 0$ can be deduced 
from it by reflection about $h=0$. It is convenient to scale out 
the parameter $J$, and quote all results in terms of the two dimensionless 
quantities $\jp /J$ and $h/J$. We will only consider an open boundary 
condition in the rung direction, namely, the summation over $a$ in the 
first term of (\ref{ham4}) runs over $1,2$. 

\begin{figure}[htb]
\begin{center}
\epsfig{figure=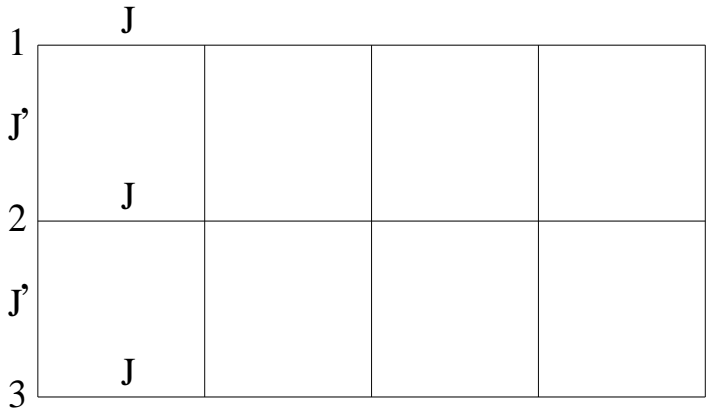,height=6cm,width=10cm}
\end{center}
\caption{Schematic picture of the three-chain ladder described in Eq. 
(\ref{ham4}). The labels $1$, $2$ and $3$ denote the three chains.}
\label{fig8}
\end{figure}

We now discuss the LEH approach for studying 
the properties of spin ladders. There are two possible limits which may be 
considered. One could examine $\jp /J \rightarrow 0$ which corresponds to 
weakly interacting chains, and then directly use techniques from bosonization 
and conformal field theory; this has been done in detail by others
\cite{cabr,tots1}. We therefore consider the strong-coupling
limit $J /\jp \rightarrow 0$ which corresponds to almost decoupled rungs. In
that limit, the LEH has been derived to first order in $J /\jp$ for a 
three-chain ladder with periodic boundary condition along 
the rungs \cite{schu2,kawa}, and for a two-chain ladder \cite{tots1,mila}. 

We derive the LEH as follows. We first set the intrachain coupling $J=0$ and 
consider which of the states of a single rung are degenerate in energy in 
the presence of a magnetic field. In general, there will be several values 
of the field, denoted by $h_0$, for which two or more of the rung states will 
be degenerate ground states. We will consider each such value of $h_0$ in 
turn. The degenerate rung states will constitute our low-energy states. If 
the degeneracy in each rung is $d$, the total
number of low-energy states in a system with $L$ rungs is given by $L^d$.
(In general, the number $d$ depends both on the system and on the 
field $h_0$. It is two for both the models we will study here).
Next, we decompose the Hamiltonian of the total system 
as $H= H_0 + V$, where $H_0$ contains only the rung interaction $\jp$ and
the field $h_0$, and $V$ contains the small interactions $J$ and the 
residual magnetic field $h - h_0$ which are both assumed to be much smaller 
than $\jp$. Let us denote the degenerate and low-energy states of the 
system as $p_i$ and the high-energy states as $q_{\alpha}$. The low-energy
states all have energy $E_0$, while the high-energy states have energies 
$E_{\alpha}$ according to the exactly solvable Hamiltonian $H_0$. Then
the first-order LEH is given, up to an additive constant, by degenerate 
perturbation theory, 
\beq
H_{eff}^{(1)} ~=~ \sum_{ij} ~\vert p_i \rangle ~\langle p_i \vert V \vert p_j 
\rangle ~\langle p_j \vert ~.
\label{ham5}
\eeq
The calculation of the various matrix elements in Eqs. (\ref{ham5}) 
can be simplified by using the symmetries of the 
perturbation $V$, e.g., translations and rotations about the $\hat z$ axis.

To derive the LEH for the three-chain ladder,
we decompose the Hamiltonian in (\ref{ham4}) as $H= H_0 + V$, where
\bea
H_0 ~&=&~ \jp ~\sum_{a=1,2} ~\sum_n ~ {\vec S}_{a,n} \cdot {\vec S}_{a+1,n} ~
-~ h_0 ~\sum_{a=1}^3 ~\sum_n ~S_{a,n}^z ~, \nonu \\
V ~&=&~ J ~\sum_{a=1}^3 ~\sum_n ~{\vec S}_{a,n} \cdot {\vec S}_{a,n+1} ~-~ 
(h ~-~ h_0 ) ~\sum_{a=1}^3 ~\sum_n ~S_{a,n}^z ~.
\label{ham6}
\eea
We determine the field $h_0$ by considering the rung Hamiltonian $H_0$ and 
identifying the values of the magnetic field $h_0$ where two or more of 
the rung states become degenerate. 

The eight states in each rung are described by specifying the $S^z$ 
components ($+$ and $-$ denoting $+1/2$ and $-1/2$ respectively) of the sites 
belonging to chains $1$, $2$ and $3$. For instance, the four states with 
total $S=3/2$ are denoted by $\vert 1 \rangle$, ..., $\vert 4 \rangle$, where 
$\vert 1 \rangle = \vert + + + \rangle$ and the other three states can be
obtained by acting on it successively with the operator $S^- = \sum_a S_a^-$.
These four states have the energy $\jp /2$ in the absence of a magnetic field. 
There is one doublet of states $\vert 5 \rangle$ and $\vert 6 \rangle$ 
with $S=1/2$, where $\vert 5 \rangle = [~2 ~\vert + - + \rangle - \vert - + + 
\rangle - \vert + + - \rangle ~]/ {\sqrt 6}$ and $\vert 6 \rangle \sim S^- 
\vert 5 \rangle$. These have energy $- \jp$. Finally, there is another 
doublet of states $\vert 7 \rangle = [~ \vert + + - \rangle - \vert - + + 
\rangle ~]/ {\sqrt 2}$ and $\vert 8 \rangle \sim S^- \vert 7 \rangle$ which 
have zero energy. It is now evident that the state $\vert 1 \rangle$
with $S^z = 3/2$ and the state $\vert 5 \rangle$ with $S^z =1/2$
become degenerate at a magnetic field $h_0 = 3\jp /2$,
while states $\vert 5 \rangle$ and $\vert 6 \rangle$ are trivially degenerate 
for the field $h_0 = 0$. We now examine these two cases separately.

For $h_0 = 3\jp /2$, the low-energy states in each rung are given by $\vert 
1 \rangle$ and $\vert 5 \rangle$, while the other six are high-energy states.
We thus have an effective spin-$1/2$ object on each rung $n$. We introduce 
three spin-$1/2$ operators $(S_n^x , S_n^y , S_n^z )$ for each rung such that
$S_n^{\pm} = S_n^x \pm i S_n^y$ and $S_n^z$ have the following actions:
\bea
S_n^+ ~\vert 1 \rangle_n ~&=&~ 0 ~, ~~~~ 
S_n^+ ~\vert 5 \rangle_n ~=~ \vert 1 \rangle_n ~, \nonu \\
S_n^- ~\vert 1 \rangle_n ~&=&~ \vert 5 \rangle_n ~, ~~~~
S_n^- ~\vert 5 \rangle_n ~=~ 0 ~, \nonu \\
{\rm and} \quad S_n^z ~\vert 1 \rangle_n ~&=&~ \frac{1}{2} ~\vert 1 
\rangle_n ~, ~~~~ S_n^z ~\vert 5 \rangle_n ~=~ -~\frac{1}{2} ~\vert 5 
\rangle_n ~. 
\label{sn}
\eea
Note that the state which has a $\vert 1 \rangle$ on every rung, i.e., 
$\vert 111 \cdots \rangle$, is just the state with rung 
magnetization $m_s =3/2$ corresponding to the saturation plateau. The state 
with a $\vert 5 \rangle$ on every rung corresponds to
the $m_s =1/2$ magnetization plateau. The LEH we are trying to derive will
therefore describe the transition between these two plateaus.

We now turn on the perturbation $V$ in (\ref{ham6}) with the assumption that 
$J$ and $h - h_0$ are both much smaller than $\jp$. We can write $V = \sum_n
V_{n,n+1}$, where 
\beq
V_{n,n+1} ~=~ J ~\sum_{a=1}^3 ~{\vec S}_{a,n} \cdot {\vec S}_{a,n+1} ~-~ 
\frac{1}{2} (h ~-~ h_0 ) ~\sum_{a=1}^3 ~[~ S_{a,n}^z ~+~ S_{a,n+1}^z ~]~.
\eeq
The action of $V_{n,n+1}$ on the four low-energy states involving rungs 
$n$ and $n+1$ can be obtained after a long but straightforward calculation.
We then use Eq. (\ref{ham5}) and find that the LEH to first order 
in $J/ \jp$ is given, up to a constant, by
\bea
H_{eff} ~=~ & & J ~\sum_n ~[~ S_n^x S_{n+1}^x ~+~ S_n^y S_{n+1}^y ~+~ 
\frac{1}{2} ~ S_n^z S_{n+1}^z ~] \nonu \\
& & -~(~ h ~-~ \frac{3\jp}{2} ~-~ \frac{J}{2} ~)~\sum_n ~S_n^z ~,
\label{ham7}
\eea
where we have substituted $h_0 = 3\jp /2$. Thus the LEH up to this order 
is simply the $XXZ$ model with anisotropy $\Delta =1/2$ in a magnetic 
field $h - 3\jp /2 - J/2$; see Eq. (\ref{ham3}). 

We now use (\ref{ham7}) to compute the values of the fields $h_1$ and
$h_2$ where the states with all rungs equal to $\vert 1 \rangle$ and all 
rungs equal to $\vert 5 \rangle$ respectively become the ground states. 
We can then identify $h_1$ with the lower critical field $h_{c-}$ for the 
plateau at $m_s =3/2$, and $h_2$ with the upper critical field $h_{c+}$ for 
the plateau at $m_s =1/2$. 

To compute the field $h_1$, we compare the energy $E_0$ of the state with 
all rungs equal to $\vert 1 \rangle $ with the minimum energy $E_{min} 
(k)$ of a spin-wave state in which one rung is equal to $\vert 5 \rangle$ 
and all the other rungs are equal to $\vert 1 \rangle$. A spin wave with 
momentum $k$ is given by
\beq
\vert k \rangle ~=~ \frac{1}{\sqrt L} ~\sum_n ~e^{ikn} ~\vert 5_n \rangle ~,
\eeq
where $\vert 5_n \rangle$ denotes a state where only the rung $n$ is 
equal to $\vert 5 \rangle$. 
The spin-wave dispersion, i.e., $\omega (k) = E (k) - E_0$, is found from 
(\ref{ham7}) to be
\beq
\omega (k) ~=~ J ~(~ \cos k ~-~ \frac{1}{2} ~)~ 
+ ~(~ h ~-~ \frac{3\jp}{2} ~-~ \frac{J}{2} ~) ~.
\eeq
This is minimum at $k =\pi$ and it turns negative there for $h < h_1$, where
\beq
h_1 ~=~ \frac{3\jp}{2} ~+~ 2 J ~.
\label{h1}
\eeq
This is therefore the transition point between the ferromagnetic state $\vert 
111 \cdots \rangle$ and a spin-wave band lying immediately below it in energy.

Similarly, we compute the field $h_2$ by comparing the energy $E_0$ of the
state with all rungs equal to $\vert 5 \rangle$ with the minimum energy 
$E_{min} (k)$ of a spin wave in which a $\vert 5 \rangle$ at one 
rung is replaced by a $\vert 1 \rangle$. For a spin wave with momentum $k$,
the dispersion $\omega (k) = E (k) - E_0$ is found to be
\beq
\omega (k) ~=~ J ~(~\cos k ~-~ \frac{1}{2} ~)~
+ ~\frac{J^2}{\jp} ~(~ \frac{2}{9} ~-~ \frac{5}{18} ~\cos 2k ~) ~
- ~(~ h ~-~ \frac{3\jp}{2} ~-~ \frac{J}{2} ~) ~.
\eeq
This is minimum at $k =\pi$ and it turns positive there for $h > h_2$, where
\beq
h_2 ~=~ \frac{3\jp}{2} ~-~ J ~.
\label{h2}
\eeq
This marks the transition between the state $\vert 555 \cdots \rangle$ and the 
spin-wave band. Equation (\ref{h2}) agrees to this order with the higher-order 
series given in the literature \cite{cabr}. 

{}From the first-order terms in (\ref{ham7}), we can deduce the asymptotic
form of the two-spin correlations. From (\ref{corrxxz}), we see that the 
exponent $\eta = 2/3$ for $\Delta =1/2$. Although this is the exponent for 
the $+-$ correlation of the effective spin-$1/2$ defined on each rung, we 
would expect the same exponent to appear in all the correlations $\langle 
S_{a,l}^+ S_{b,n}^- \rangle$ studied by DMRG in the previous section, 
regardless of how we choose the chain indices $a,b = 1,2,3$. We find that 
the analytically predicted exponent of $2/3$ agrees quite well with the 
numerically obtained exponents which lie in the range $0.61$ to $0.70$ 
\cite{tand}.

We now consider the LEH at the other magnetic field $h_0 =0$ where the
rung states $\vert 5 \rangle$ and $\vert 6 \rangle$ are degenerate. 
We take these as the low-energy states and introduce
new effective spin-$1/2$ operators for each rung with actions similar
to Eqs. (\ref{sn}), except that we replace $\vert 1 \rangle$ and $\vert 5 
\rangle$ in those equations by $\vert 5 \rangle$ and $\vert 6 \rangle$. 
We compute the action of the perturbation $V$ on the low-energy states, and 
deduce the LEH to be
\beq
H_{eff} ~=~ J ~\sum_n ~{\vec S}_n \cdot {\vec S}_{n+1} ~-~ h ~\sum_n ~S_n^z ~.
\label{ham8}
\eeq
This Hamiltonian describes the transition between the magnetization plateaus 
at $m_s =1/2$ and $m_s =-1/2$; since these plateaus are reflections of each
other about zero magnetic field, it is sufficient to study one of them. By a 
calculation similar to the one used to derive (\ref{h1}), the field $h_1$ 
can be found from the dispersion of a spin wave in which one rung
is equal to $\vert 6 \rangle$ and all the other rungs are equal to $\vert 5
\rangle$. The dispersion is
\beq
\omega (k) ~=~ h ~+~ J ~(~ \cos k ~-~ 1 ~)~~.
\eeq
This gives
\beq
h_1 ~=~ 2J ~.
\eeq
This is the lower critical field $h_{c-}$ of the $m_s =1/2$ plateau. 
The Hamiltonian (\ref{ham8}) describes an isotropic
spin-$1/2$ antiferromagnet. From the comments made earlier, we see 
that this model only has the two saturation plateaus at $m_s =\pm 1/2$, and no 
other plateau in between. For $h=0$, the two-spin correlations decay as power 
laws with the exponent $\eta =1$; see Eq. (\ref{corrxxz}).

We now use the LEH approach to study a two-chain spin-$1/2$
ladder with the following Hamiltonian,
\bea
H ~=~ & & \jp ~\sum_n ~ {\vec S}_{1,n} \cdot {\vec S}_{2,n} ~+~ J_2 ~
\sum_{a=1}^2 ~\sum_n ~{\vec S}_{a,n} \cdot {\vec S}_{a,n+1} \nonu \\ 
& & +~ 2 J_1 ~\sum_n ~{\vec S}_{1,n} \cdot {\vec S}_{2,n+1} ~-~ h ~
\sum_{a=1}^2 ~\sum_n ~S_{a,n}^z ~,
\label{ham9}
\eea
as shown in Fig. 9. The model may be viewed as a single chain with an 
alternation in nearest-neighbor couplings $\jp$ and $2J_1$ (dimerization), and 
a next-nearest-neighbor coupling $J_2$ (frustration). Eq. (\ref{ham9}) has 
been studied from the point of view of magnetization plateaus using a 
first-order LEH, bosonization and exact diagonalization 
\cite{tots1,tone,mila}. 

\begin{figure}[htb]
\begin{center}
\epsfig{figure=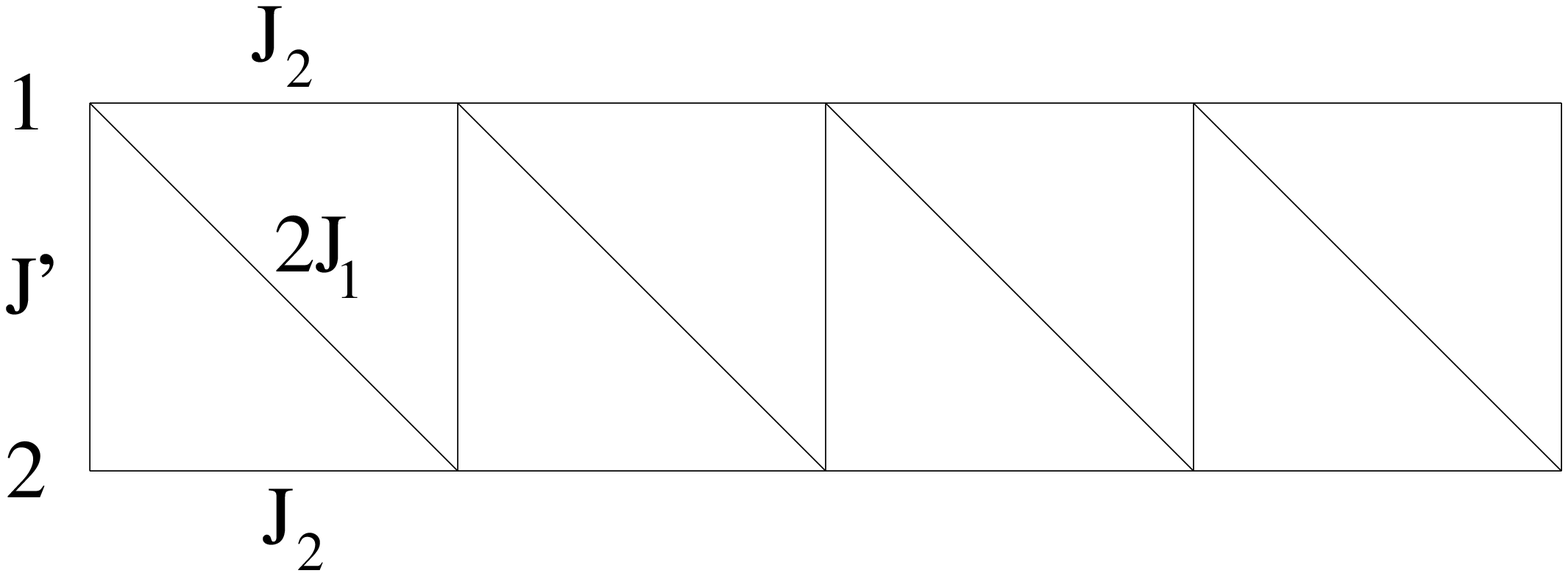,height=4cm,width=10cm}
\end{center}
\caption{Schematic picture of the two-chain ladder described in Eq. 
(\ref{ham9}). The labels $1$ and $2$ denote the two chains.}
\label{fig9}
\end{figure}

We begin by setting $J_1 = J_2 =0$, and studying the four states on each rung.
These are specified by giving the configurations $\pm$ of the spins on
chains $1$ and $2$ as follows. The three triplet states with $S=1$ are
denoted as $\vert 1 \rangle$, $\vert 2 \rangle$ and $\vert 3 \rangle$, where
$\vert 1 \rangle = \vert + + \rangle$ and the other two states are obtained
by acting on it successively with $S^-$. These three states
have energy $\jp /4$ in the absence of a magnetic field. The singlet state
$\vert 4 \rangle = [ \vert + - \rangle - \vert - + \rangle ]/ {\sqrt 2}$
has energy $-3\jp /4$. The states $\vert 1 \rangle$ and $\vert 4 \rangle$ 
become degenerate at a field $h_0 = \jp$. We now develop perturbation
theory by assuming that $J_1 , J_2$ and $h - h_0$ are all much less than
$\jp$. The perturbation is $V = \sum_n V_{n,n+1}$ where
\bea
V_{n,n+1} ~=~ & & J_2 ~\sum_{a=1}^2 ~ {\vec S}_{a,n} \cdot {\vec S}_{a,n+1} ~
+~ 2 J_1 ~{\vec S}_{1,n} \cdot {\vec S}_{2,n+1} \nonu \\
& & -~\frac{1}{2} (h ~-~ h_0 ) ~\sum_{a=1}^2 ~[~S_{a,n}^z ~+~ S_{a,n+1}^z ~]~.
\eea
The actions of this operator on the four low-energy states of a pair of 
neighboring rungs can be easily obtained. We now introduce effective 
spin-$1/2$ operators ${\vec S}_n$ on each rung which act on the two low-energy 
states. The LEH is then found to be
\bea
H_{eff} ~= & & (~ J_2 - J_1 ~) ~\sum_n ( S_n^x S_{n+1}^x + S_n^y S_{n+1}^y ) ~
+ ~\frac{1}{2} (~ J_2 + J_1 ~) ~\sum_n S_n^z S_{n+1}^z \nonu \\
& & -~ (~ h ~-~ \jp ~-~ \frac{J_1}{2} ~-~ \frac{J_2}{2} ~) ~\sum_n ~S_n^z ~.
\label{ham10}
\eea

We now compute the field $h_1$ above which the state $\vert 111 \cdots
\rangle$ becomes the ground state. The dispersion of a spin wave, in which 
one rung is equal to $\vert 4 \rangle$ and all the others are equal to $\vert
1 \rangle$, is given by 
\beq
\omega (k) ~=~ h ~-~ \jp ~-~ J_1 ~-~ J_2 ~+~ (J_2 - J_1 ) ~\cos k ~.
\eeq
By minimizing this as a function of $k$ in various regions in 
the parameter space $(J_1 , J_2 )$, and then setting that minimum value equal
to zero, we find that $h_1$ is given by 
\bea
h_1 ~&=&~ \jp ~+~ 2 J_1 ~~{\rm if}~~ J_2 \le J_1 ~, \nonu \\
&=&~ \jp ~+~ 2 J_2 ~~{\rm if}~~ J_2 \ge J_1 ~.
\eea
This is the lower critical field $h_{c-}$ of the saturation plateau with
magnetization $m_s =1$ per rung. Similarly, we can find the field $h_2$ from 
the dispersion of a spin wave in which one rung is equal to $\vert 1 \rangle$ 
and the rest are equal to $\vert 4 \rangle$. The dispersion is given by 
\beq
\omega (k) ~=~ -~ h ~+~ \jp ~+~ (J_2 - J_1 ) ~\cos k ~.
\eeq
By setting the minimum of this equal to zero, we find that $h_2$ is given by 
\bea
h_2 ~&=&~ \jp ~+~ J_2 ~-~ J_1 ~~ {\rm if} ~~ J_2 \le J_1 ~, \nonu \\
&=&~ \jp ~-~ J_2 ~+~ J_1 ~~ {\rm if} ~~ J_2 \ge J_1 ~.
\eea
This is the upper critical field $h_{c+}$ of the saturation plateau with
magnetization $m_s =0$ per rung.

Finally, we can see that the first-order terms in (\ref{ham10}) are of the 
same form as the $XXZ$ model in (\ref{ham3}). We can always make the
coefficient of the first term in (\ref{ham10}) positive, if necessary by 
performing a rotation $S_n^x \rightarrow (-1)^n S_n^x$, $S_n^y \rightarrow 
(-1)^n S_n^y$ and $S_n^z \rightarrow S_n^z$. We then get a Hamiltonian of the 
form
\bea
H_{eff} ~=~ & & \vert J_2 ~-~ J_1 \vert ~\sum_n ~[~ S_n^x S_{n+1}^x ~+~ 
S_n^y S_{n+1}^y ~]~ +~ \frac{1}{2} ~(~ J_2 ~+~ J_1 ~)~ \sum_n ~S_n^z 
S_{n+1}^z \nonu \\
& & -~ (~ h ~-~ \jp ~-~ \frac{J_1}{2} ~-~ \frac{J_2}{2} ~)~ \sum_n ~S_n^z ~.
\label{ham11}
\eea
This is an $XXZ$ model with 
\beq
\Delta ~=~ \frac{J_2 ~+~ J_1}{2 ~\vert J_2 ~-~ J_1 \vert} ~.
\eeq
{}From the earlier comments, we see that the two-chain ladder 
has an additional plateau at $m_s =1/2$ for $\Delta > 1$, i.e., if $J_2 
+ J_1 > 2 \vert J_2 - J_1 \vert$. In particular, $\Delta = \infty$ for $J_2 
= J_1$; the $m_s =1/2$ plateau should then extend all the way from the upper 
critical field of the $m_s =0$ plateau to the lower critical field of the 
$m_s =1$ plateau. This can be seen in Fig. 10 which is taken from Ref. 
\cite{tone}; the dimerization parameter $\alpha$ in that figure is related to 
our couplings by $\jp = 1 + \alpha$ and $2J_1 = 1 - \alpha$. Note that the 
$m_s =1/2$ plateau is particularly broad at $\alpha = 0.6$, i.e., $J_2 = J_1 
= 0.2$, and that it actually touches the $m_s =1$ plateau on the right. The 
fact that it does not extend all the way up to the $m_s =0$ plateau on the 
left is probably because we have ignored higher-order terms which lead to 
deviations from the $XXZ$ model.

\begin{figure}[htb]
\begin{center}
\epsfig{figure=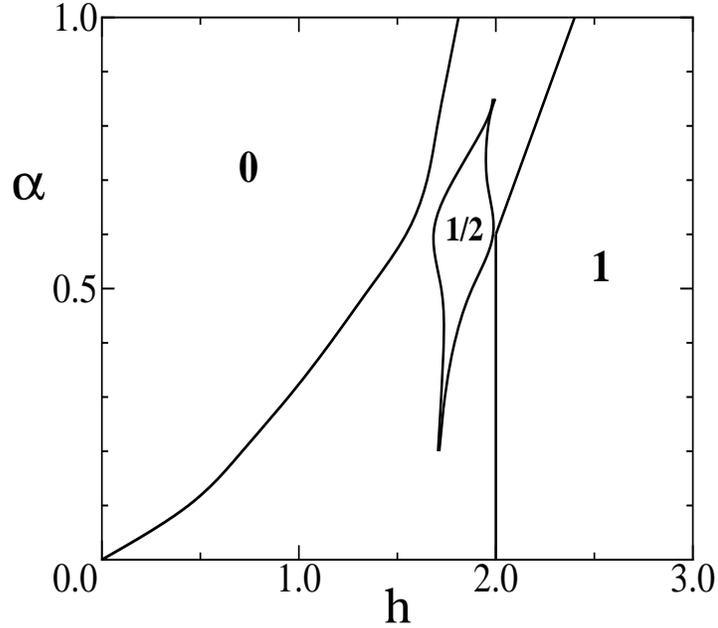,height=9cm,width=10cm}
\end{center}
\vspace*{-0.2cm}
\caption{Magnetization plateaus of the two-chain ladder as a function $h$ and
$\alpha$ for $J_2 = 0.2$. The numbers $0$, $1/2$ and $1$ correspond to the
values of $m_s$ at the plateaus.}
\label{fig10}
\end{figure}

To summarize,
we studied a three-chain spin-$1/2$ ladder with a large ratio of interchain
coupling to intrachain coupling using a LEH approach. We found a wide plateau
with rung magnetization given by $m_s =1/2$. The two-spin correlations are 
extremely short-ranged in the plateau. All these are consistent with the 
large magnetic gap. At other values of $m$, the two-spin correlations fall 
off as power laws; the exponents can be found by using the first-order LEH 
which takes the form of an $XXZ$ model in a longitudinal magnetic field. 
We also used the LEH approach to study a two-chain 
ladder with an additional diagonal interaction. In addition to a plateau at 
$m_s =0$, this system also has a plateau at $m_s =1/2$ for certain regions in 
parameter space. The $m_s =1/2$ plateau is interesting because it corresponds
to degenerate ground states which spontaneously break the translation
invariance of the Hamiltonian. This can be understood from the LEH which, at
first-order, is an $XXZ$ model with $\Delta > 1$.

\section{Summary}

We have presented some field theoretic methods for studying the properties
of quantum spin systems in one dimension. Each of these methods has
a particular regime of validity (i.e., large $S$ for the NLSMs, small
$S$ for bosonization, and weak perturbations for the LEHs) within which 
the method can give a reasonable qualitative picture of the ground state 
and low-energy excitations. Such a picture is very useful for gaining
a quick understanding of a given model, even though one may then need
to use numerical methods like the DMRG to obtain quantitative results. 

\vskip 1 true cm


\begin{thebibliography}{99}

\bibitem{hald1} F. D. M. Haldane, Phys. Lett. {\bf 93A}, 464 (1983); Phys.
Rev. Lett. {\bf 50}, 1153 (1983).

\bibitem{buye} W. J. L. Buyers, R. M. Morra, R. L. Armstrong, M. J. Hogan, P. 
Gerlach and K. Hirakawa, Phys. Rev. Lett. {\bf 56}, 371 (1986); J. P. Renard, 
M. Verdaguer, L. P. Regnault, W. A. C. Erkelens, J. Rossat-Mignod and W. G. 
Stirling, Europhys. Lett. {\bf 3}, 945 (1987); S. Ma, C. Broholm, D. H. Reich, 
B. J. Sternlieb and R. W. Erwin, Phys. Rev. Lett. {\bf 69}, 3571 (1992).

\bibitem{dago} E. Dagotto and T. M. Rice, Science {\bf 271}, 618 (1996).

\bibitem{eccl} R. S. Eccleston, T. Barnes, J. Brody and J. W. Johnson, Phys. 
Rev. Lett. {\bf 73}, 2626 (1994).

\bibitem{azum} M. Azuma, Z. Hiroi, M. Takano, K. Ishida and Y. Kitaoka, Phys. 
Rev. Lett. {\bf 73}, 3463 (1994).

\bibitem{chab1} G. Chaboussant, P. A. Crowell, L. P. Levy, O. Piovesana, A.
Madouri and D. Mailly, Phys. Rev. B {\bf 55}, 3046 (1997).

\bibitem{hase} M. Hase, I. Terasaki and K. Uchinokura, Phys. Rev. Lett. {\bf 
70}, 3651 (1993); M. Nishi, O. Fujita and J. Akimitsu, Phys. Rev. B {\bf 50}, 
6508 (1994); G. Castilla, S. Chakravarty and V. J. Emery, Phys. Rev. Lett. 
{\bf 75}, 1823 (1995).

\bibitem{chab2} G. Chaboussant, Y. Fagot-Revurat, M.-H. Julien, M. E. Hanson,
C. Berthier, M. Horvatic, L. P. Levy and O. Piovesana, Phys. Rev. Lett. {\bf
80}, 2713 (1998).

\bibitem{oshi} M. Oshikawa, M. Yamanaka and I. Affleck, Phys. Rev. Lett. {\bf
78}, 1984 (1997).

\bibitem{cabr} D. C. Cabra, A. Honecker and P. Pujol, Phys. Rev. B {\bf 58},
6241 (1998). See also the web site 
http://thew02.physik.uni-bonn.de/$\sim$honecker/roc.html

\bibitem{tots1} K. Totsuka, Phys. Rev. B {\bf 57}, 3454 (1998).

\bibitem{tone} T. Tonegawa, T. Nishida and M. Kaburagi, preprint no. 
cond-mat/9712297, to appear in Physica B (Proc. $5^{th}$ Int. Symp. on 
Research in High Magnetic Fields, Sydney, 1997).

\bibitem{saka} K. Sakai and M. Takahashi, Phys. Rev. B {\bf 57}, R3201 (1998); 
K. Sakai and M. Takahashi, Phys. Rev. B {\bf 57}, R8091 (1998).

\bibitem{schu1} H. J. Schulz, G. Cuniberti and P. Pieri, in {\it Field 
Theories for Low Dimensional Condensed Matter Systems}, edited by G. Morandi, 
A. Tagliacozzo and P. Sodano (Springer, Berlin, 2000), cond-mat/9807366; H. 
J. Schulz, in {\it Proceedings of Les Houches Summer School LXI}, edited by E. 
Akkermans, G. Montambaux, J. Pichard and J. Zinn-Justin (Elsevier, Amsterdam, 
1995), cond-mat/9503150.

\bibitem{affl1} I. Affleck, in {\it Fields, Strings and Critical Phenomena}, 
eds. E. Brezin and J. Zinn-Justin (North-Holland, Amsterdam, 1989); I. Affleck 
and F.D.M. Haldane, Phys. Rev. B {\bf 36}, 5291 (1987).

\bibitem{reig} M. Reigrotzki, H. Tsunetsugu and T. M. Rice, J. Phys. Condens. 
Matter {\bf 6}, 9235 (1994); T. Barnes, J. Riera and D. A. Tennant, Phys. Rev.
B {\bf 59}, 11384 (1999).

\bibitem{frad} E. Fradkin, {\it Field Teories of Condensed Matter Systems} 
(Addison-Wesley, Reading, 1991).

\bibitem{whit} S. R. White, Phys. Rev. Lett. {\bf 69}, 2863 (1992); Phys. Rev. 
B {\bf 48}, 10345 (1993).

\bibitem{pati} S. Pati, R. Chitra, D. Sen, S. Ramasesha, and H. R. 
Krishnamurthy, J. Phys. Condens. Matter {\bf 9}, 219 (1997). 

\bibitem{chit} R. Chitra, S. K. Pati, H. R. Krishnamurthy, D. Sen and S. 
Ramasesha, Phys. Rev. B {\bf 52}, 6581 (1995).

\bibitem{hamm} P. R. Hammar and D. H. Reich, J. Appl. Phys. {\bf 79}, 5392
(1996); C. A. Hayward, D. Poilblanc and L. P. Levy, Phys. Rev. B {\bf 54},
R12649 (1996).

\bibitem{shas} B. S. Shastry and B. Sutherland, Phys. Rev. Lett. {\bf 47}, 964 
(1981).

\bibitem{rao1} S. Rao and D. Sen, J. Phys. Condens. Matter {\bf 9}, 1831 
(1997).

\bibitem{affl2} I. Affleck, Nucl. Phys. B {\bf 265}, 409 (1986).

\bibitem{tots2} K. Totsuka, Y. Nishiyama, N. Hatano, and M. Suzuki, J. Phys. 
Condens. Matter {\bf 7}, 4895 (1995); Y. Kato and A. Tanaka, J. Phys. Soc. 
Jpn. {\bf 63}, 1277 (1994).

\bibitem{rao2} S. Rao and D. Sen, Nucl. Phys. B {\bf 424}, 547 (1994).

\bibitem{alle} D. Allen and D. Senechal, Phys. Rev. B {\bf 51}, 6394 (1995). 

\bibitem{suth} B. Sutherland, Phys. Rev. B {\bf 12}, 3795 (1975).

\bibitem{norm} B. Normand, J. Kyriakidis and D. Loss, Ann. Phys. (Leipzig)
{\bf 9}, 133 (2000).

\bibitem{gogo} A. O. Gogolin, A. A. Nersesyan and A. M. Tsvelik, {\it 
Bosonization and Strongly Correlated Systems} (Cambridge University Press, 
Cambridge, 1998).

\bibitem{shan} R. Shankar, Lectures given at the BCSPIN School, Kathmandu, 
1991, in {\it Condensed Matter and Particle Physics}, edited by Y. Lu, J. Pati 
and Q. Shafi (World Scientific, Singapore, 1993); J. von Delft and H. 
Schoeller, Annalen der Physik {\bf 7}, 225 (1998), cond-mat/9805275; S. Rao 
and D. Sen, cond-mat/0005492.

\bibitem{hald2} F. D. M. Haldane, Phys. Rev. Lett. {\bf 45}, 1358 (1980);
Phys. Rev. Lett. {\bf 47}, 1840 (1981); J. Phys. C {\bf 14}, 2585 (1981).

\bibitem{okam} K. Okamoto and K. Nomura, Phys. Lett. A {\bf 169}, 433 (1992).

\bibitem{mila} F. Mila, Eur. Phys. J. B {\bf 6}, 201 (1998); A. K. Kolezhuk, 
Phys. Rev. B {\bf 59}, 4181 (1999).

\bibitem{schu2} H. J. Schulz, in {\it Strongly Correlated Magnetic and 
Superconducting Systems}, edited by G. Sierra and M. A. Martin-Delgado,
Lecture Notes in Physics 478 (Springer, Berlin, 1997).

\bibitem{kawa} K. Kawano and M. Takahashi, J. Phys. Soc. Jpn. {\bf 66}, 4001 
(1997).

\bibitem{tand} K. Tandon, S. Lal, S. K. Pati, S. Ramasesha and D. Sen, Phys. 
Rev. B {\bf 59}, 396 (1999).

\end{thebibliography}
\end{document}